\pdfoutput=1

\documentclass[a4paper]{PoS}

\title{Lattice QCD at Non-Zero Temperature}

\ShortTitle{Lattice QCD at Non-Zero Temperature}

\author{\speaker{Alexei Bazavov}$^{a,b}$\\
\llap{$^a$}Department of Physics and Astronomy, University of California, Riverside, CA 92521, USA\\
\llap{$^b$}Department of Physics and Astronomy, University of Iowa, Iowa City, IA 52242, USA\\
        E-mail: \email{obazavov@quark.phy.bnl.gov}
}

\abstract{
This is a review of selected recent developments in finite-temperature
lattice QCD. The focus is on the properties of the chiral crossover region,
deconfinement and fluctuations of conserved charges, the equation of state,
properties of heavy quarkonia and reconstruction of spectral functions.
}

\FullConference{The 32nd International Symposium on Lattice Field Theory\\
		 23-28 June, 2014\\
		 Columbia University, New York, NY}

\begin{document}

\section{Introduction}

Shortly after the discovery of asymptotic 
freedom~\cite{Gross:1973id,Politzer:1973fx} in
non-Abelian gauge theories~\cite{Yang:1954ek}, existence of a new phase of matter
was suggested at large baryon density~\cite{Collins:1974ky} or high
temperature \cite{Cabibbo:1975ig}.
The transition to this new state of matter, quark-gluon plasma (QGP),
happens at typical hadronic scales, $O(\Lambda_{QCD})$, where the 
strong-coupling constant, $\alpha_s$ is $O(1)$, and, thus,
weak-coupling expansions are not suitable to address this problem.

Wilson's non-perturbative regularization of quantum field theories
on a space-time lattice~\cite{Wilson:1974sk} provided the necessary framework
to study gauge theories at strong coupling, and after pioneering
numerical studies of the $SU(2)$ gauge theory by Creutz~\cite{Creutz:1980zw},
investigations of the finite-temperature transition and
properties of the deconfined phase
followed~\cite{McLerran:1980pk,Kuti:1980gh,Engels:1980ty}.

Experimentally quark-gluon plasma
can be achieved by colliding heavy nuclei at high enough energies.
This experimental program is being carried out at the Relativistic Heavy-Ion
Collider at BNL and Large Hadron Collider (LHC) at CERN. One of the unexpected
findings of RHIC was that the matter created in
the collisions behaves not as a weakly coupled gas, but as a strongly
coupled fluid~\cite{Arsene:2004fa,Back:2004je,Adams:2005dq,Adcox:2004mh}.

\section{Chiral crossover}

In the path-integral representation the partition function of QCD on the
lattice can be written as:
\begin{equation}
Z=\int DUD\bar\psi D\psi e^{-S[U,\bar\psi,\psi]},\,\,\,\,\,
S[U,\bar\psi,\psi]=S_g[U]+S_f[U,\bar\psi,\psi],
\end{equation}
where $S_g$ is the gauge and $S_f$ is the fermion action.
Path integrals of this form can be
evaluated approximately by designing a Markov Chain Monte Carlo
process and sampling the most probable field configurations.

\begin{figure}[h]
\begin{center}
\includegraphics[width=0.42\textwidth]{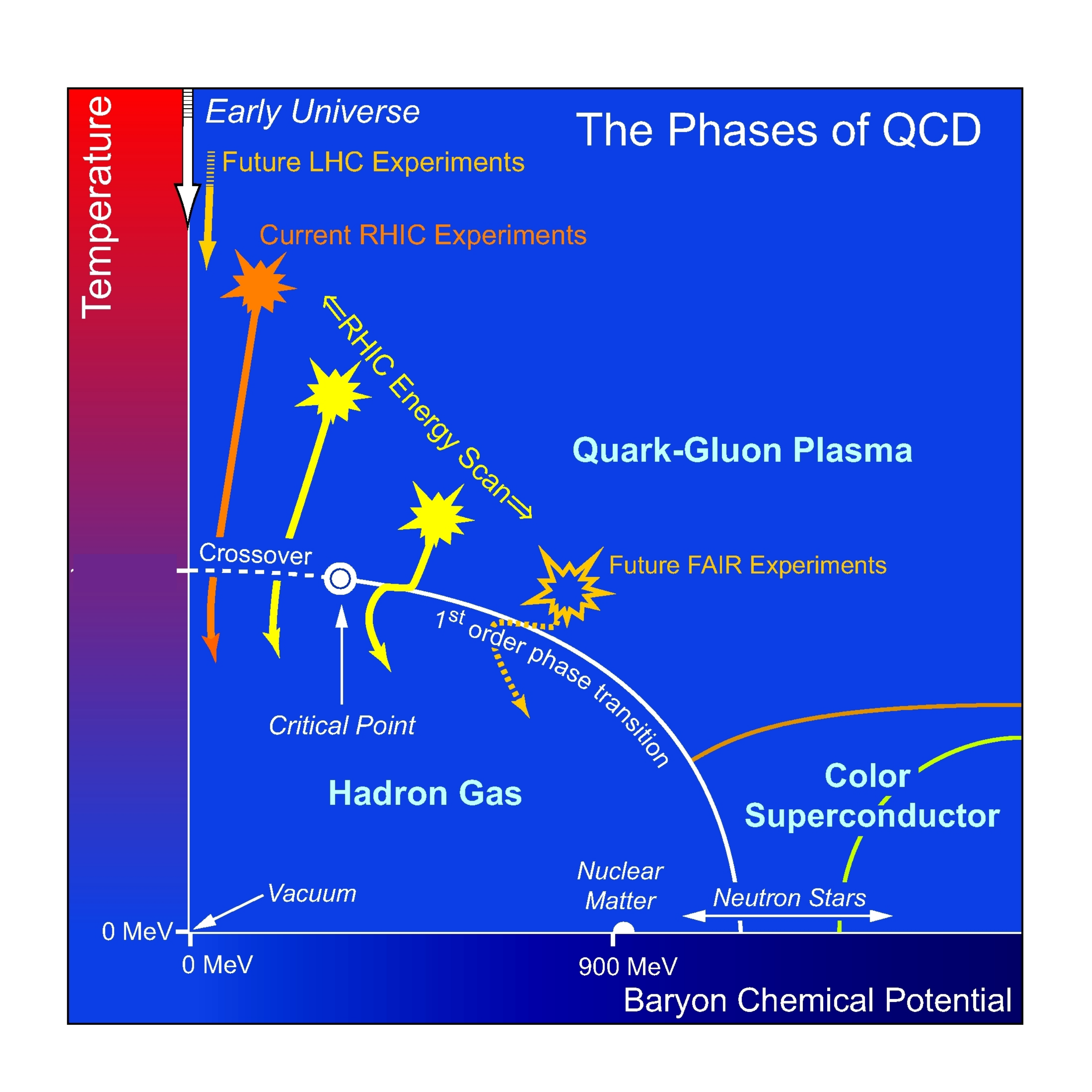}\hfill
\includegraphics[width=0.462\textwidth]{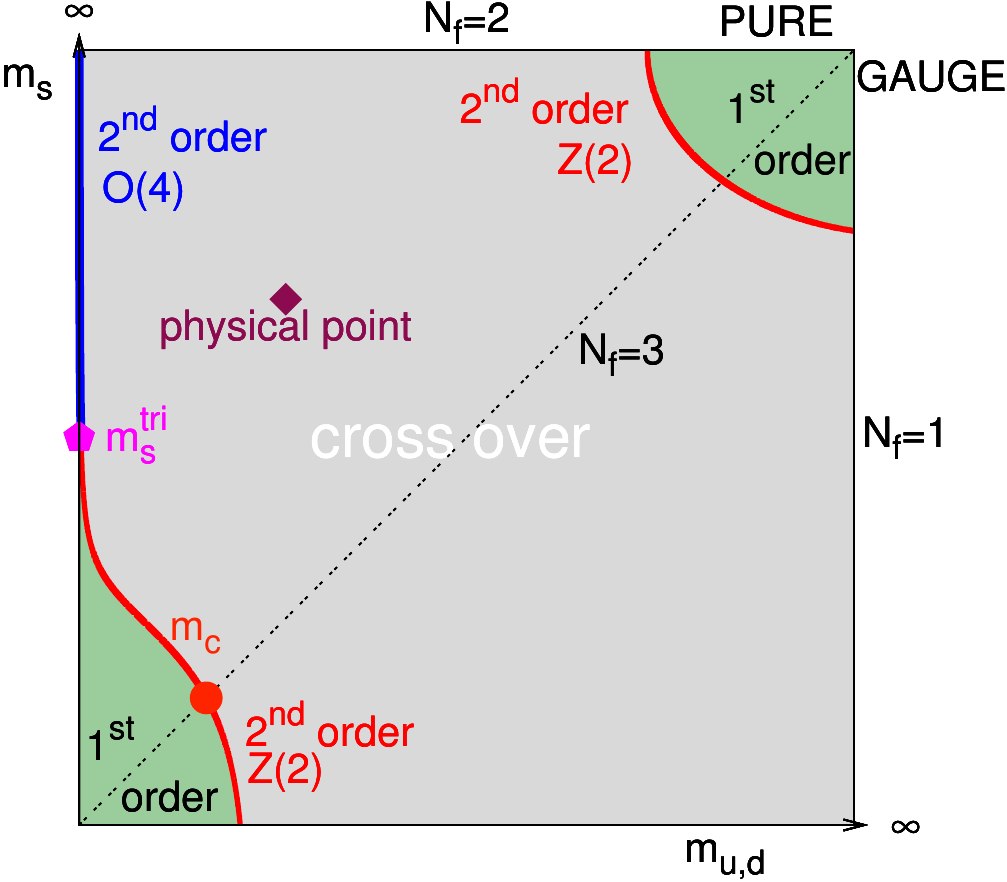}
\parbox[b]{0.45\textwidth}{
\caption{
The phase diagram of QCD in the $\mu_B-T$ plane.
}
\label{fig_QCD_phases}
}
\hfill
\parbox[b]{0.46\textwidth}{
\caption{
The phase diagram of QCD in the $m_l-m_s$ plane at $\mu_B=0$, from~\cite{Ding:2015ona}.
}
\label{fig_columbia}
}
\end{center}
\end{figure}

It has been established by lattice calculations that at the physical
values of light and strange quark masses at zero baryon chemical
potential the transition from the hadronic into the deconfined phase
of QCD is a rapid crossover rather than a genuine phase
transition~\cite{Bernard:2004je,Cheng:2006qk,Aoki:2006we}.
The conjectured phase diagram of QCD in $\mu_B-T$ plane is shown
in Fig.~\ref{fig_QCD_phases}. At low temperature and small baryon
density the strongly interacting matter is in the hadronic phase,
where quarks are confined in hadrons and the $SU(2)_R\times SU(2)_L$
symmetry (of the massless Lagrangian) is spontaneously broken
by the vacuum. At high temperature and/or large baryon chemical potential,
quarks and gluons are deconfined and the chiral symmetry is restored.

Direct Monte Carlo simulations are possible
along $\mu_B=0$ axis, and the region of not too large $\mu_B/T$ is
accessible indirectly through Taylor expansions in $\mu_B/T$ or
reweighting techniques. Thus, at the moment, the location of the line
of first order phase transitions, the critical point, and nature
of the phases at large $\mu_B$ are open questions yet to be
answered by \textit{ab initio} calculations. Interesting ideas are being
pursued to extend Monte Carlo simulations into the region of non-zero
chemical potential~\cite{Sexty:2014dxa}.

However, even at $\mu_B=0$ QCD possesses certain critical behavior,
depending on the fermion content of the theory. The phase diagram
of QCD in $m_l(=m_u=m_d)-m_s$ plane (the Columbia plot~\cite{Brown:1990ev})
is shown in Fig.~\ref{fig_columbia}. Studies with staggered fermions
put the physical point in the crossover region, not too far
from the two-flavor chiral limit $m_l\to0$, which is expected to be in the $O(4)$
universality class. (This can be quantified by studying how well
the universal scaling behavior with corrections due to finite $m_l$
can be applied to the chiral condensate $\langle\bar\psi\psi\rangle$,
which is the order parameter in the chiral limit, and its susceptibility
$\chi=\partial \langle\bar\psi\psi\rangle/\partial m_l$.)

\begin{figure}[b]
\begin{center}
\includegraphics[width=0.455\textwidth]{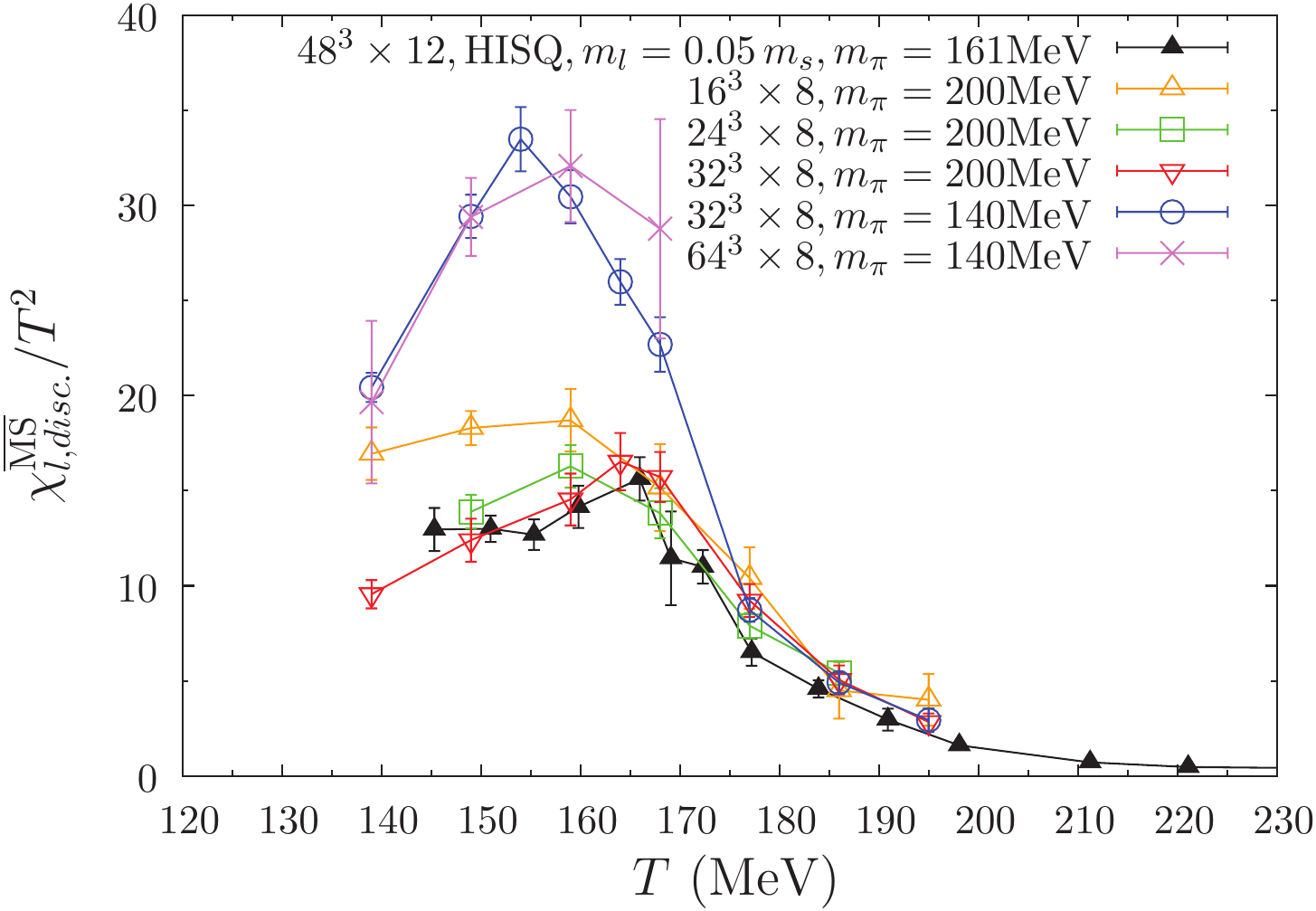}\hfill
\includegraphics[width=0.50\textwidth]{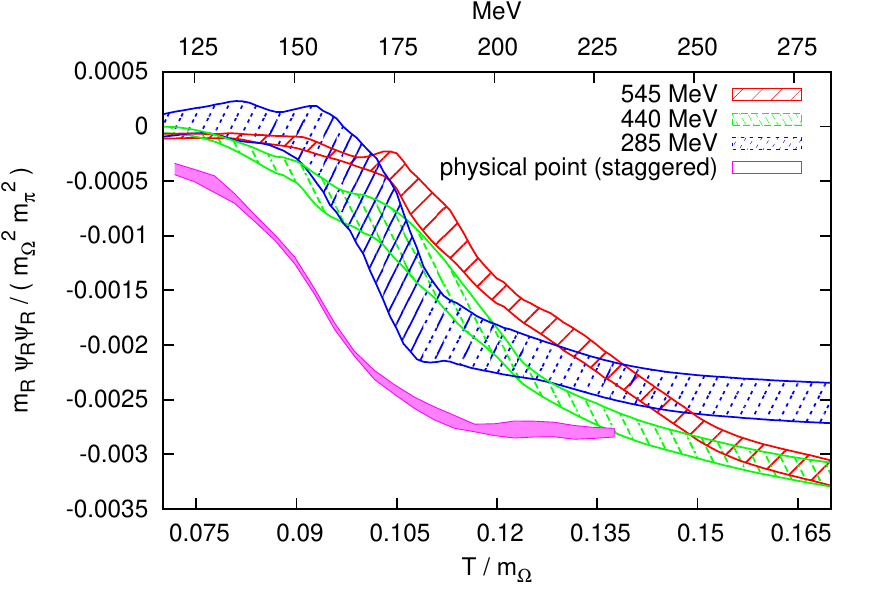}
\parbox[b]{0.45\textwidth}{
\caption{
Disconnected chiral susceptibility with DWF at $m_\pi=135$ and 
$200$~MeV~\cite{Bhattacharya:2014ara}, compared with HISQ~\cite{Bazavov:2011nk}.
}
\label{fig_chi_dwf}
}
\hfill
\parbox[b]{0.46\textwidth}{
\caption{
The renormalized chiral condensate with Wilson fermions at three values
of the pion mass, in the continuum limit~\cite{Borsanyi:2015waa}.
}
\label{fig_pbp_wilson}
}
\end{center}
\end{figure}

The chiral crossover temperature has been determined in the
continuum limit at the physical values of light quark masses by the
Wuppertal-Budapest ($T_c=147(2)(3)$~MeV defined from the peak in the chiral
susceptibility, $T_c=157(3)(3)$ or $155(3)(3)$~MeV from the inflection
point of the chiral condensate renormalized in two different
ways)~\cite{Aoki:2006we,Aoki:2009sc,Borsanyi:2010bp}
and HotQCD ($T_c=154(8)(1)$~MeV from the $O(4)$ scaling analysis of the
chiral condensate and susceptibility)~\cite{Bazavov:2011nk}
collaborations using staggered fermions. Given the large cutoff effects due
to violations of the taste symmetry of staggered fermions at computationally
feasible lattice spacings, it required a use
of improved fermionic actions, such as stout~\cite{Morningstar:2003gk} and
highly improved staggered quarks (HISQ)~\cite{Follana:2006rc},
respectively, to reliably calculate this quantity.

It is also important to crosscheck the staggered result in calculations with
other types of lattice fermions. The HotQCD collaboration has continued studying
the transition region with the domain-wall fermions
(DWF)~\cite{Kaplan:1992bt,Furman:1994ky}, which are
in particular well-suitable for capturing the chiral aspects of QCD.
Simulations~\cite{Bhattacharya:2014ara}
were performed directly at the physical pion mass, $m_\pi=135$~MeV,
but due to the high computational cost, only at one lattice cutoff $N_\tau=8$.
The chiral crossover temperature has been determined from the location of the peak
in the disconnected chiral susceptibility, shown in Fig.~\ref{fig_chi_dwf}.
The end result, including the systematic uncertainty is $T_c=155(1)(8)$~MeV, in agreement
with the staggered value~\cite{Bazavov:2011nk}.

The Wuppertal-Budapest collaboration continued finite-temperature calculations
with Wilson fermions~\cite{Borsanyi:2015waa} at several values of the pion mass,
$m_\pi=285$, $440$ and $545$~MeV. The results for the renormalized chiral
condensate are shown in Fig.~\ref{fig_pbp_wilson} together with the staggered
result at the physical pion mass. They show the correct trend of shifting the 
transition region to lower temperatures with decreasing the pion mass, however,
the uncertainties are still too large for a quantitative comparison and require
further study.

\section{Restoration of the axial symmetry}

\begin{figure}
\centering
\includegraphics[width=0.47\textwidth]{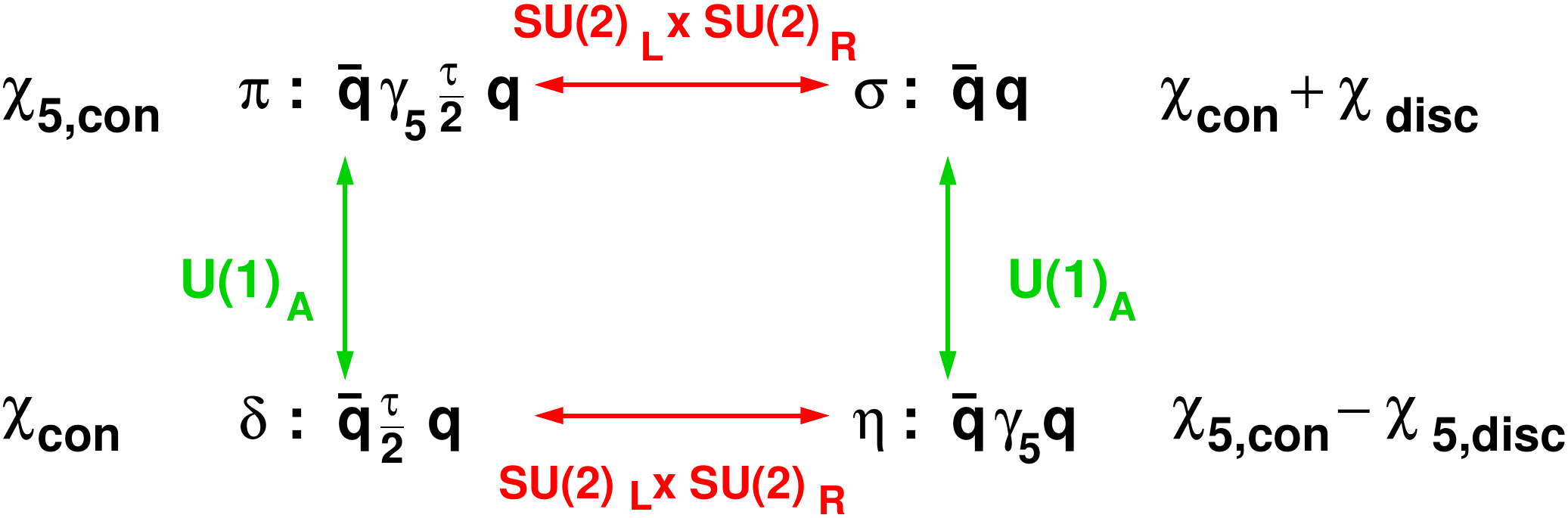}
\caption{
Symmetry transformations relating scalar and pseudo-scalar mesons,
from~\cite{Bazavov:2012qja}.
}
\label{fig_chiral_U1A}
\end{figure}

\begin{figure}[b]
\begin{center}
\includegraphics[width=0.47\textwidth]{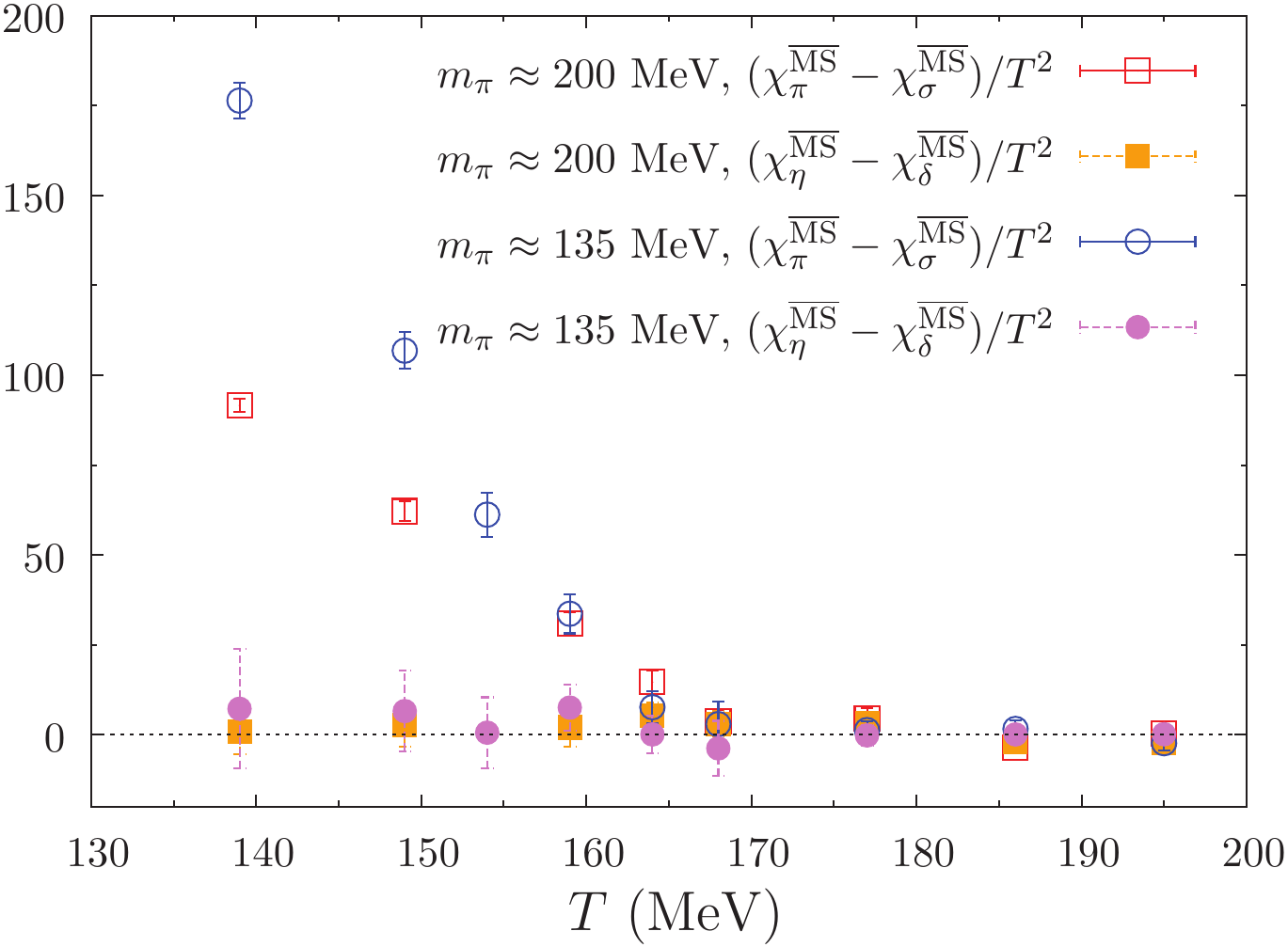}\hfill
\includegraphics[width=0.47\textwidth]{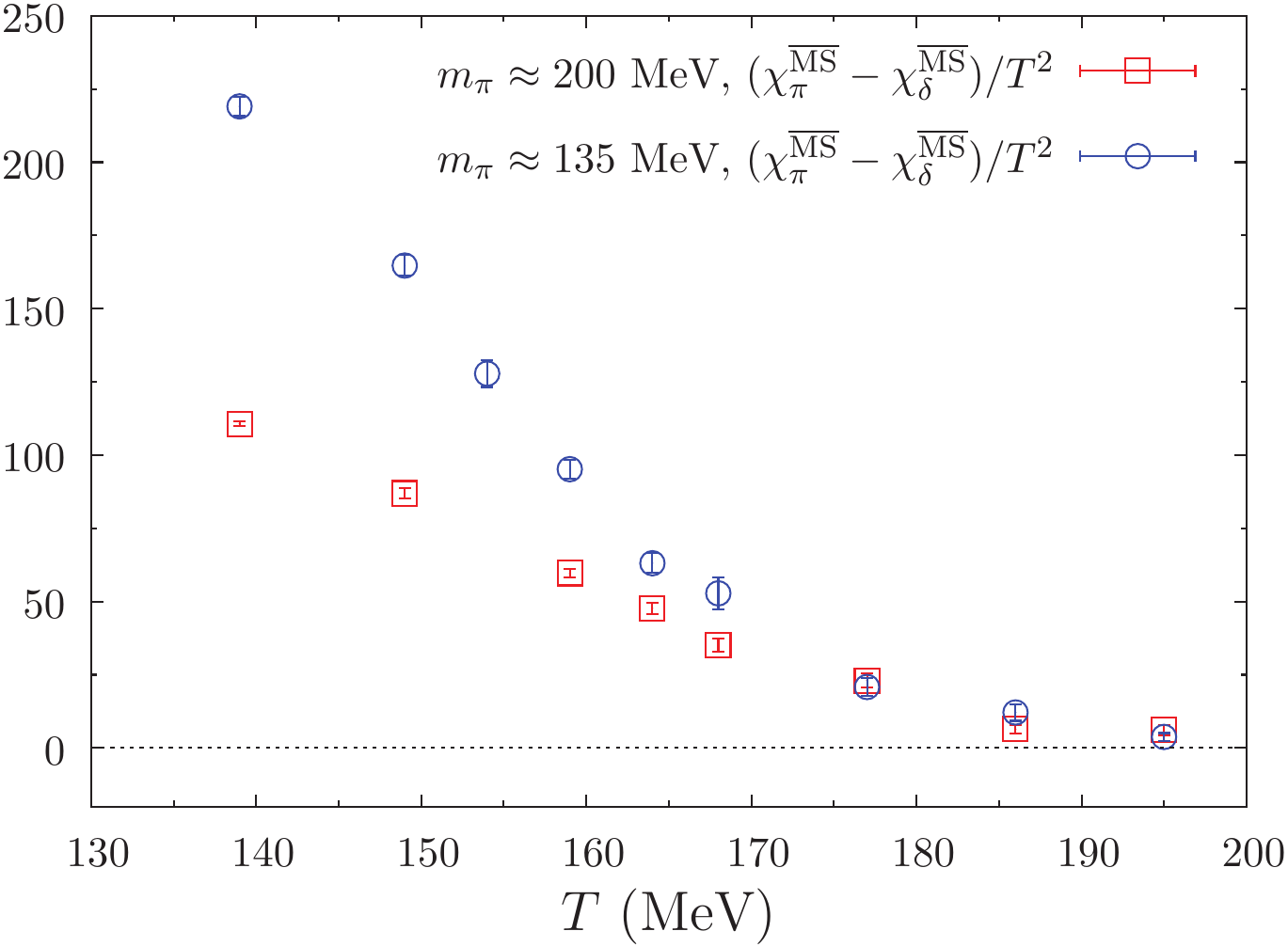}
\caption{
The chiral, $\chi_\pi-\chi_\sigma$ and
$\chi_\eta-\chi_\delta$ (left), and the axial, $\chi_\pi-\chi_\delta$ (right),
symmetry-breaking differences with domain-wall fermions~\cite{Bhattacharya:2014ara}.
}
\label{fig_chiral_dwf}
\end{center}
\end{figure}

While lattice simulations with staggered fermions suggest that the transition in
2+1 flavor QCD in the limit of vanishing light quark masses, $m_l\to0$, is of second
order and belongs to the $O(4)$ universality class, the nature of this transition is far
from settled. This is due to the effect of the axial anomaly on the finite-temperature
transition.
The full symmetry group of the
classical massless Lagrangian is $SU(2)_L\times SU(2)_R\times U_V(1)\times U_A(1)$,
however the axial $U_A(1)$ symmetry is anomalous and is broken on the quantum
level~\cite{Adler:1969gk,Bell:1969ts}.
If the $U_A(1)$ symmetry remains significantly broken around the chiral transition,
then $O(4)$ universality class is appropriate, but if it is effectively restored, the
transition may be of first order~\cite{Pisarski:1983ms}
or second order, but in a different universality
class~\cite{Pelissetto:2013hqa}. The strength of symmetry breaking can be analyzed
from the behavior of susceptibilities, which are the 4-volume integrals of
the two-point correlation functions in various channels, \textit{e.g.}
$\chi_\pi=\int d^4x\langle \pi(0)\pi(x)\rangle$ for the pion. Restoration of a symmetry
would lead to degeneracy in the spectrum, and the symmetry transformations relating
pseudo-scalar and scalar mesons are shown in Fig.~\ref{fig_chiral_U1A}.

The HotQCD collaboration reported~\cite{Bhattacharya:2014ara}
results on the chiral symmetry-breaking differences $\chi_\pi-\chi_\sigma$
and $\chi_\eta-\chi_\delta$, Fig.~\ref{fig_chiral_dwf} (left) and $U_A(1)$-breaking
difference $\chi_\pi-\chi_\delta$, Fig~\ref{fig_chiral_dwf} (right) with domain-wall
fermions at the physical pion mass. While $\chi_\pi-\chi_\sigma$ is large
below $T_c$, it becomes zero for $T>164$~MeV, indicating restoration of
the chiral symmetry. On the contrary, the $U_A(1)$ difference remains large
at $T_c$, indicating no restoration of this symmetry until at least $T>196$~MeV.

The $U_A(1)$ breaking difference can be related to the spectral density $\rho(\lambda)$
of the eigenvalues of the Dirac operator as
\begin{equation}
\chi_\pi-\chi_\delta=\lim_{m\to0}\int_0^\infty d\lambda \rho(\lambda)
\frac{4m^2}{(m^2+\lambda^2)^2}.
\end{equation}
The JLQCD collaboration studied the spectrum of the Dirac operator
with the M\"{o}bius domain-wall and overlap
fermions~\cite{Cossu:2014aua,Tomiya:2014mma}.
The results seem to support their earlier findings~\cite{Aoki:2012yj}
that $\rho(\lambda)$ starts with cubic powers of $\lambda$ and when
the chiral symmetry is restored, the $\chi_\pi-\chi_\delta$ difference
becomes insensitive to the axial symmetry breaking, and that the axial
symmetry may be effectively restored close to $T_c$. Future work may
be needed to clarify the fate of the $U_A(1)$ symmetry and the
nature of the transition in the chiral limit.

\section{Fluctuations of conserved charges}

A powerful tool to explore the crossover region and the nature
of the degrees of freedom in the deconfined phase is fluctuations
and correlations of various conserved charges. We can define 
generalized susceptibilities as derivatives of the pressure:
\begin{equation}
\chi_{klmn}^{BQSC}=\left.\frac{\partial^{(k+l+m+n)}
[p(\hat\mu_B,\hat\mu_Q,\hat\mu_S,\hat\mu_C)/T^4]}
{\partial\hat\mu_B^k\partial\hat\mu_Q^l
\partial\hat\mu_S^m\partial\hat\mu_C^n}
\right|_{\mu=0}
\label{eq_gen}
\end{equation}
where $\hat\mu_i=\mu_i/T$, $i=B,Q,S,C$ are the dimensionless chemical
potentials for the baryon number, electric charge, strangeness and
charm, respectively. (We use a convention to drop a superscript
in $\chi_{klmn}^{BQSC}$ when the corresponding subscript is zero.)

At low temperature fluctuations of various charges are suppressed due
to confinement, while at high temperature they should approach
ideal quark gas values. It has been recently shown how the electric
charge fluctuations can be used for \textit{ab initio}
determination of the freeze-out parameters (\textit{i.e.}
values of the temperature, $T^f$, and baryon chemical potential,
$\mu_B^f$, at which system hadronizes at the end of the evolution
in heavy-ion collisions)~\cite{Bazavov:2012vg,Borsanyi:2013hza} and
also to probe the strangeness carrying degrees of freedom in the
high-temperature phase~\cite{Bazavov:2013dta,Bellwied:2013cta}. These
results have been summarized in the last year's review by
Szabo~\cite{Szabo:2014iqa}.

There are several new developments this year. The electric charge
fluctuations are of particular interest, because they can be measured
well in the heavy-ion experiments. Unfortunately, in calculations
with staggered fermions they are the most sensitive to the taste
breaking effects and require substantial computational effort for full
control of the continuum limit. The Wuppertal-Budapest collaboration
reported about ongoing calculations of $\chi^Q_2$ and $\chi^Q_4$ and
attempts of continuum extrapolation~\cite{Borsanyi2014_unpub}.
The BNL-Bielefeld-CCNU collaboration continued calculations of the
sixth-order cumulants with the aim of determining possible critical
behavior and exploring the QCD phase diagram~\cite{Schmidt2014_unpub}.

The phenomenology of heavy-ion collisions provides strong
evidence~\cite{BraunMunzinger:2003zd}
that the thermodynamics of the hadronic phase up to about
the chiral crossover temperature can be well described by a gas
of uncorrelated hadrons and resonances -- the Hadron Resonance Gas (HRG)
model~\cite{Hagedorn:1965st}. The HRG partition function is a product
of individual partition functions of all states in the QCD spectrum,
often approximated by taking all known states from the particle
data tables (PDG)~\cite{Beringer:1900zz}.

\begin{figure}[h]
\begin{center}
\parbox[b]{0.47\textwidth}{
\includegraphics[width=0.45\textwidth]{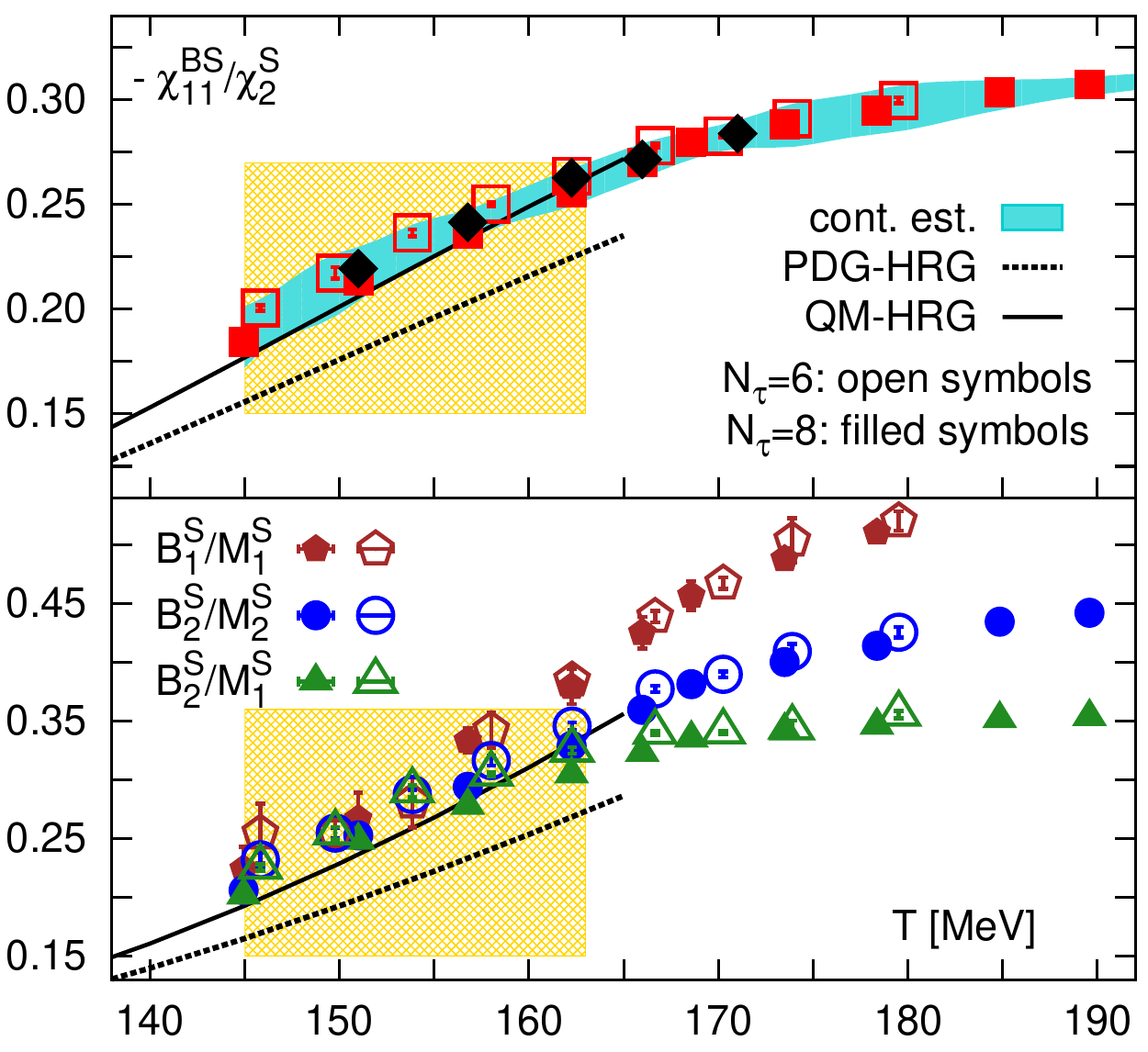}\hfill
\caption{
Ratios of susceptibilities (top) and linear combinations of
susceptibilities that project baryonic and mesonic sectors (bottom)
calculated on the lattice (symbols) and in the hadron resonance gas model
with the PDG spectrum (solid lines) and with the spectrum predicted
by the quark model (dotted lines). The HRG with the PDG spectrum significantly
underpredicts the lattice data~\cite{Bazavov:2014yba}.
\label{fig_strange_rat}
}
}
\hfill
\parbox[b]{0.47\textwidth}{
\includegraphics[width=0.45\textwidth]{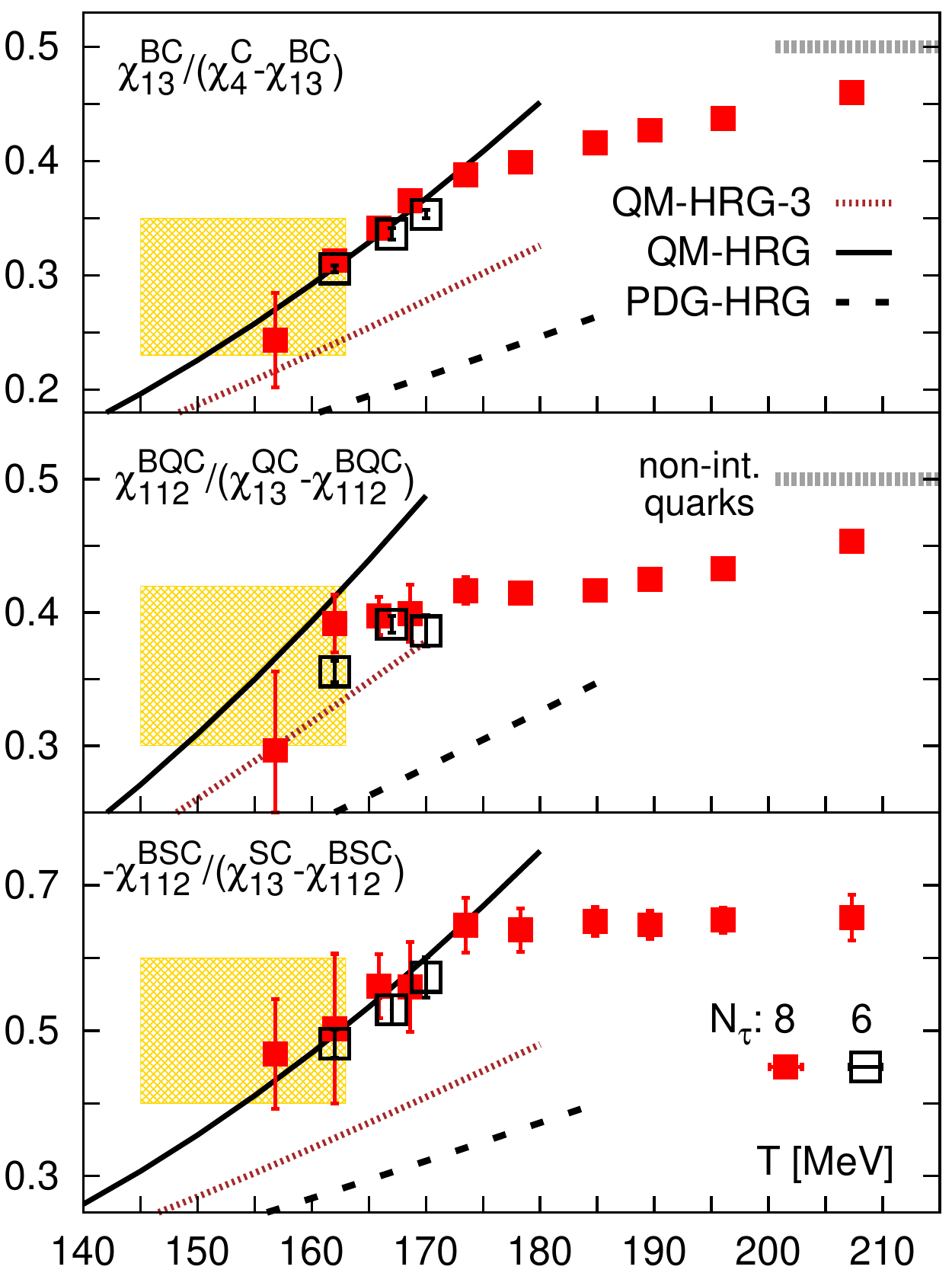}
\caption{
Similar to Fig.~{\protect\ref{fig_strange_rat}} for observables that include the charm quark~\cite{Bazavov:2014xya}.
\label{fig_charm_rat}
}
}
\end{center}
\end{figure}

It has been noted that certain quantities, such as strangeness
fluctuations and correlations of strangeness and baryon number,
are larger in QCD than in HRG with PDG spectrum~\cite{Borsanyi:2011sw,Bazavov:2012jq},
and argued that these differences provide evidence for contribution
of additional, not yet observed experimentally, hadron resonances.
The BNL-Bielefeld-CCNU collaboration reported~\cite{Ding:2014bua}
their findings~\cite{Bazavov:2014yba,Bazavov:2014xya} on the
possible thermodynamic relevance of these states.
Out of various generalized susceptibilities, Eq.~(\ref{eq_gen}),
calculated on the lattice with HISQ fermions and two cutoffs,
$N_\tau=6$ and $8$,
they construct combinations, that project on various baryon number
and strangeness sectors. The results are then compared with two versions of
the HRG model: with spectrum taken from PDG and with the one that includes states
predicted by quark model calculations, Fig.~\ref{fig_strange_rat} and
\ref{fig_charm_rat}. Clearly, including the additional states
provides a much better description of the lattice data.
This also has implications for the determination of the freeze-out
temperature for strange hadrons~\cite{Bazavov:2014xya}.

While at low temperature the susceptibilities help to test the HRG
model, and in the transition region indicate a switch in the dominant
degrees of freedom in the system, at high temperatures, deeply in the
deconfined phase, they should eventually reach perturbative behavior.
A recent calculation of the second (in the continuum) and fourth
(at several cutoffs) order quark number susceptibilities indicates
reasonable agreement with weak-coupling expansions for 
$T\gtrsim400$~MeV~\cite{Bazavov:2013uja}, given the uncertainties of the latter.
However, more work is needed for controlling the cutoff effects and taking
the continuum limit for the fourth-order susceptibilities at high
temperature.

\section{The equation of state}

A change from the confined into the deconfined phase 
is signaled by the rapid increase in the pressure, $p$, and
energy density, $\epsilon$.
These quantities signal deconfinement of the degrees of
freedom with quantum numbers of quarks and gluons and approach the
Stefan-Boltzmann limit at asymptotically high temperatures.

A lattice calculation of the equation of state usually starts with
evaluation of the trace of the energy-momentum tensor, also called the
trace anomaly, or the interaction measure:
\begin{equation}
\frac{\Theta^{\mu\mu}(T)}{T^4} =\frac{\epsilon -3p}{T^4}=T\frac{d}{d T}
\left(\frac{p}{T^4}\right).
\label{eq_theta}
\end{equation}
The pressure can then be calculated by integrating the trace anomaly,
starting from some reference value $p_0(T_0)$:
\begin{equation}
\label{eq_pT4}
\frac{p(T)}{T^4} = \frac{p_0}{T_0^4} + \int_{T_0}^T dT'\frac{\Theta^{\mu\mu}}{T^{\prime5}}.
\end{equation}
All other thermodynamic quantities that are derivatives of the partition
function with respect to the temperature can be calculated from
Eqs. (\ref{eq_theta}) and (\ref{eq_pT4}), using various thermodynamic identities.

\subsection{$N_f=0$}

What makes the calculation of the equation of state on the lattice numerically
expensive is the additive renormalization arising from the breaking of the
Lorentz symmetry by the lattice. For every value of the gauge coupling one
needs to evaluate a zero-temperature subtraction for the trace anomaly.
The fixed-scale approach can potentially reduce the cost of zero-temperature
calculations, since in this scheme the temperature $T=1/(aN_\tau)$ is varied
by changing $N_\tau$ at fixed $a$ (so only one zero-temperature calculation
is needed for several temperatures). Possible temperatures are then limited
by a few integer values of $N_\tau$.

\begin{figure}[h]
\centering
\includegraphics[width=0.445\textwidth]{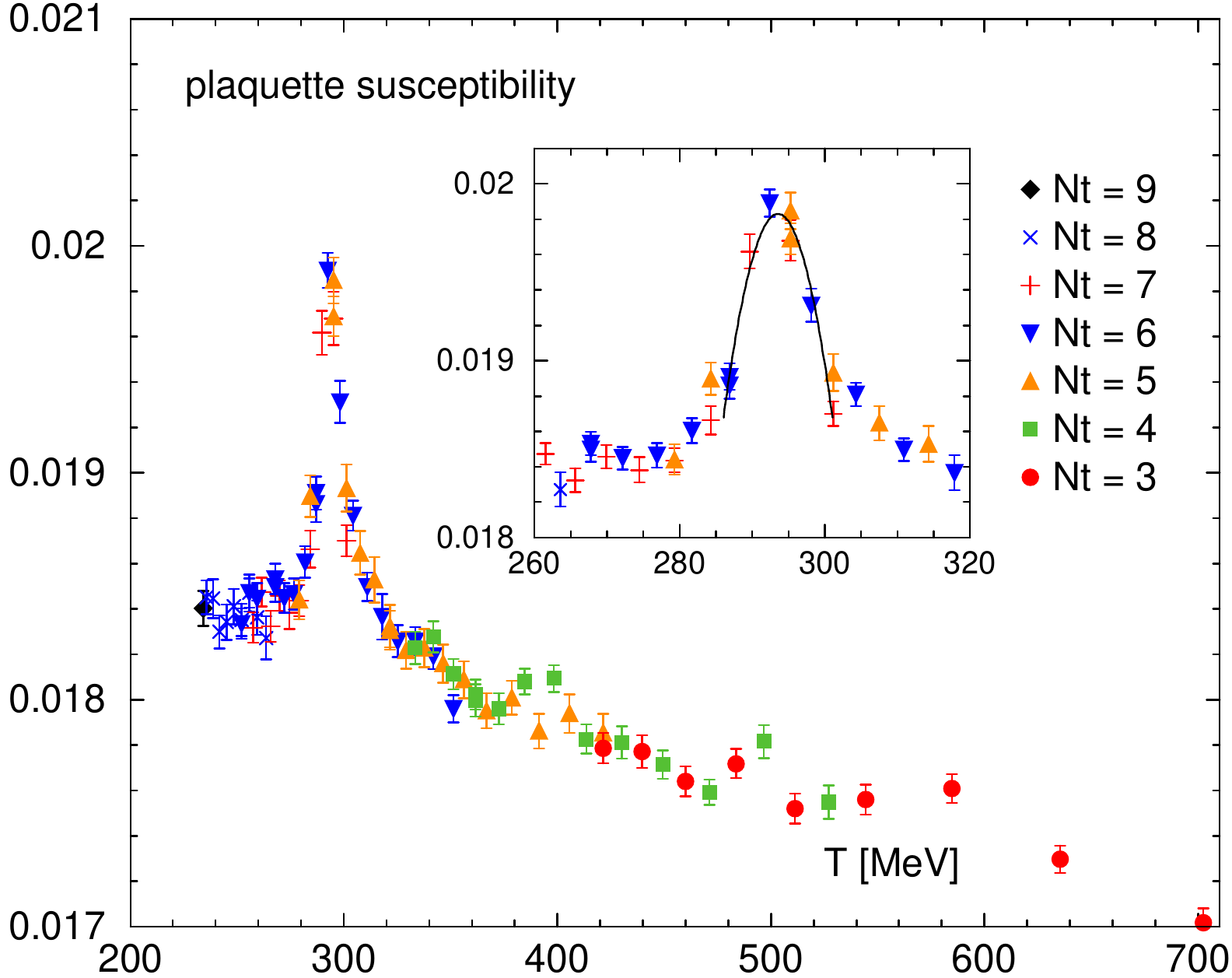}\hfill
\includegraphics[width=0.42\textwidth]{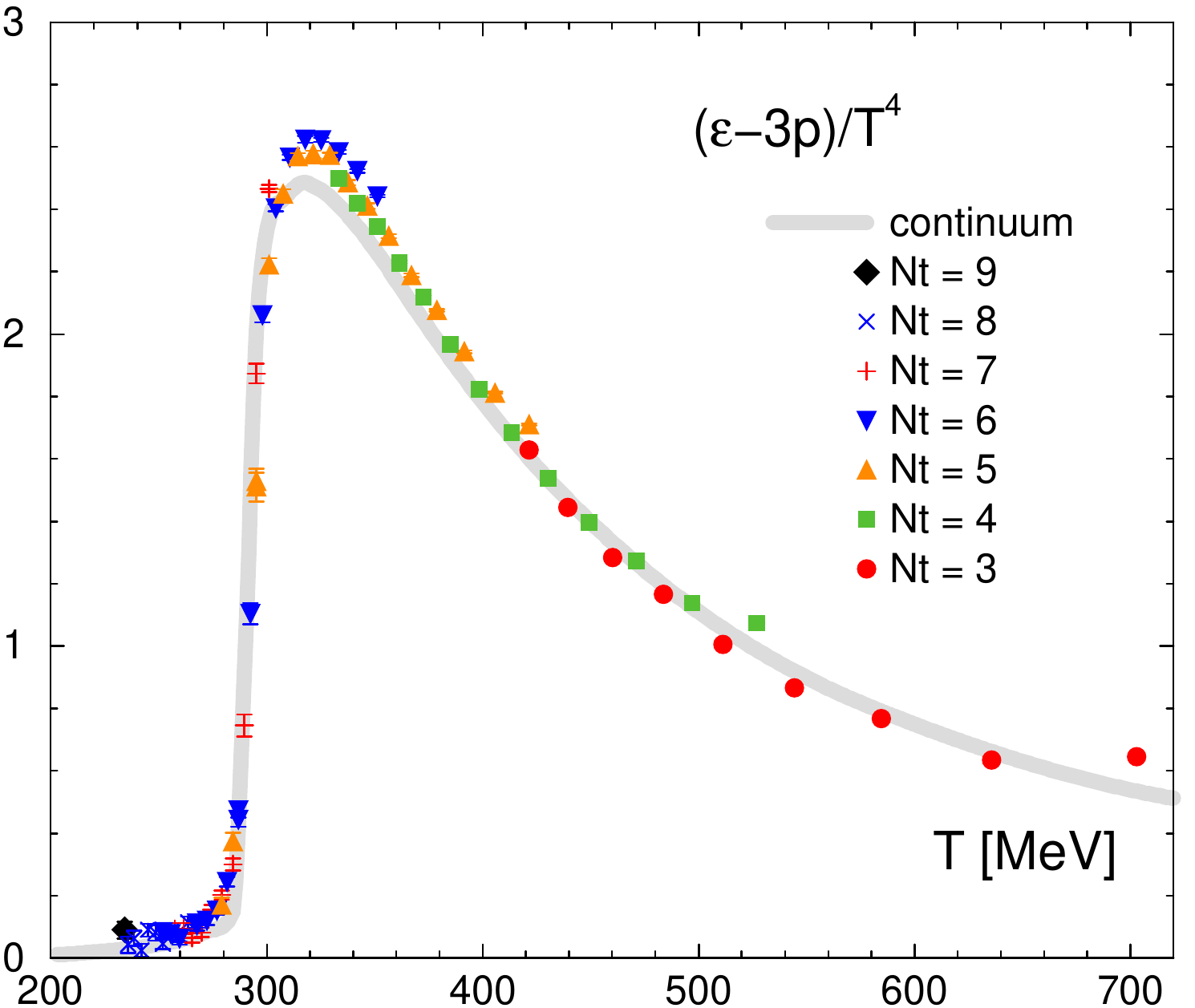}
\caption{
The plaquette susceptibility (left) and the trace anomaly (right) for $SU(3)$ pure
gauge theory, calculated at one lattice spacing using shifted boundary conditions.
}
\label{fig_eos_nf0}
\end{figure}

By using recently proposed shifted boundary conditions~\cite{Giusti:2010bb}
\begin{equation}
U_4(\vec{x},N_\tau)=U_4(\vec{x}+\vec{s},0),
\end{equation}
where $U$ is the gauge link variable, $N_\tau$ is the temporal extent of the
lattice and $\vec{s}$ is the shift vector, one can cover
more temperatures, $T=1/\sqrt{a^2N_\tau^2+\vec{s}^2}$, by varying $\vec{s}$
in conjunction with $N_\tau$. A calculation for $SU(3)$ pure gauge theory
at single lattice spacing of about $0.1$~fm has been recently reported~\cite{Umeda:2014ula}.
The plaquette susceptibility, used to determine $T_c=293$~MeV, is shown in
Fig.~\ref{fig_eos_nf0} (left) and the trace anomaly
in Fig.~\ref{fig_eos_nf0} (right) together with the continuum result
of Ref.~\cite{Borsanyi:2012ve}.

\subsection{$N_f=2$}

\begin{figure}[b]
\centering
\includegraphics[width=0.35\textwidth]{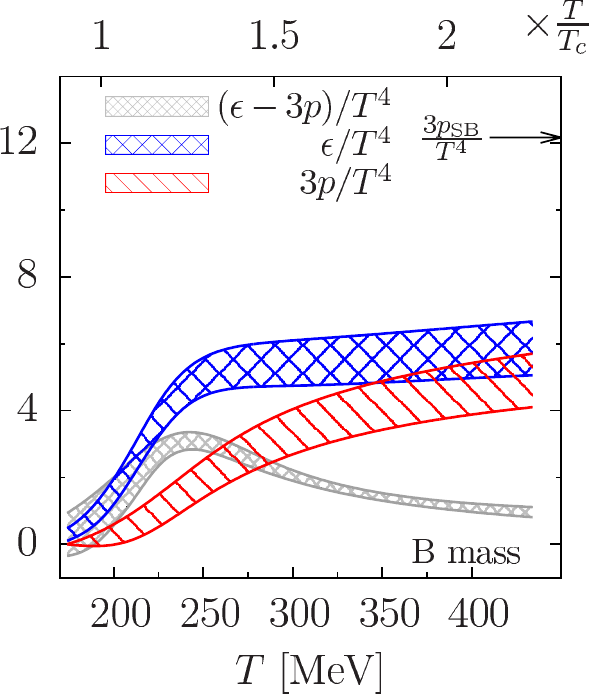}\hfill
\includegraphics[width=0.49\textwidth]{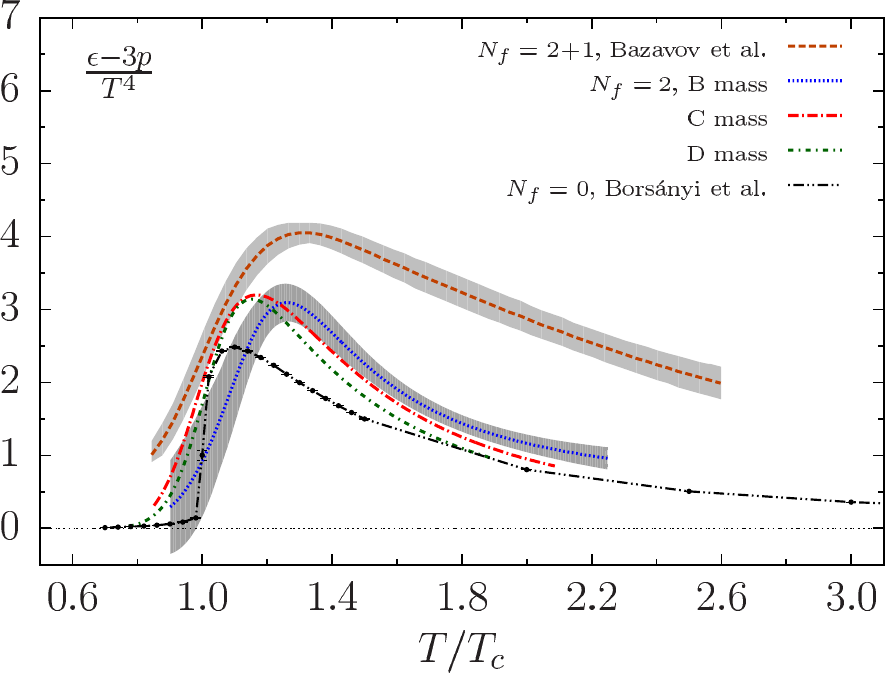}
\caption{
Left: The trace anomaly, energy density and pressure for two flavors
of twisted mass Wilson fermions at $m_\pi=360$~MeV. Right: The trace anomaly
for two flavors of twisted mass Wilson fermions at three values of $m_\pi$
compared with the pure gauge and 2+1 flavor staggered results.
}
\label{fig_eos_nf2}
\end{figure}

The tmfT collaboration made progress in calculation of the equation
of state with two flavors of twisted mass Wilson fermions~\cite{Burger:2014xga}.
The simulations were performed at three values of the pion mass,
$m_\pi=360$, $430$ and $640$~MeV and the continuum limit was taken.
The results for the trace anomaly, pressure and energy density for the lowest
pion mass are shown in Fig.~\ref{fig_eos_nf2} (left).
Around 400~MeV the energy density and pressure reach only about half
of the Stefan-Boltzmann limit value.
And in Fig.~\ref{fig_eos_nf2} (right) the two-flavor trace
anomaly for three pion masses is compared with the 
pure gauge~\cite{Borsanyi:2012ve} and
2+1 flavor~\cite{Bazavov:2014pvz} results.

\subsection{$N_f=2+1$}

\begin{figure}
\centering
\includegraphics[width=0.47\textwidth]{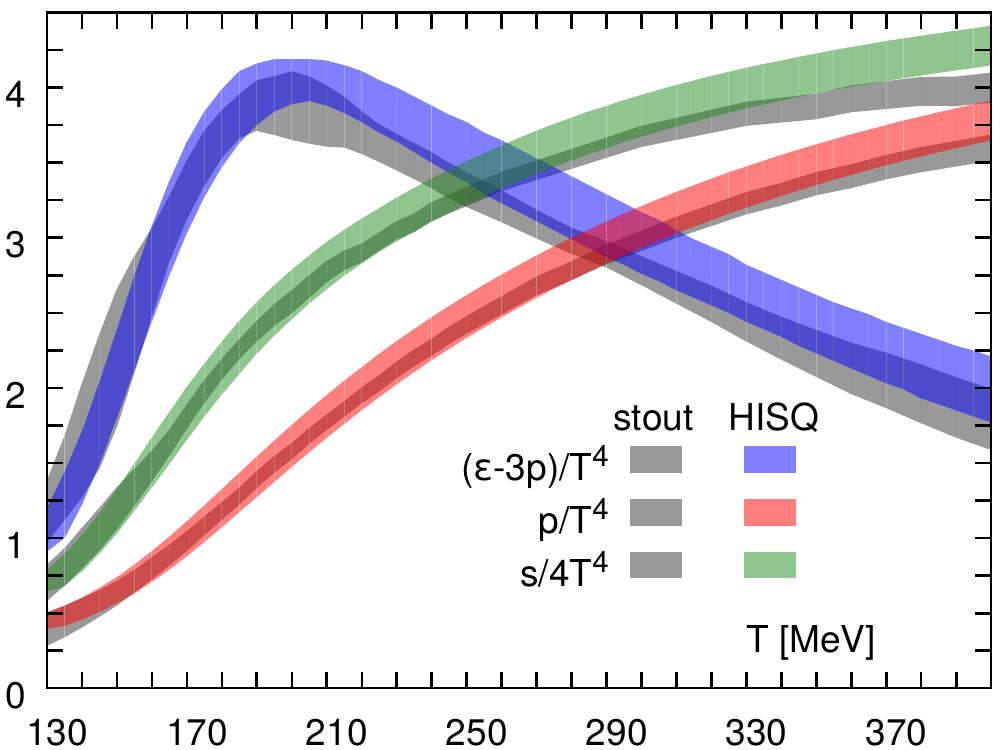}\hfill
\includegraphics[width=0.47\textwidth]{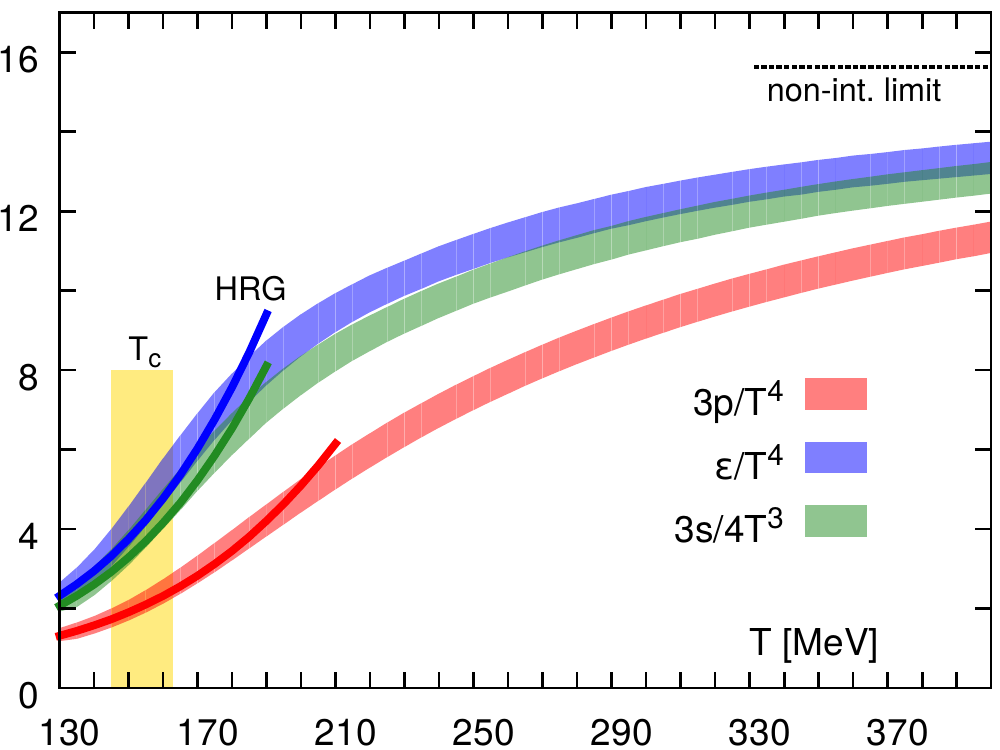}
\parbox[b]{0.46\textwidth}{
\caption{
Comparison of the continuum results for the trace anomaly, pressure
and entropy density for the stout and HISQ action.
}
\label{fig_e3p_p_s}
}
\hfill
\parbox[b]{0.46\textwidth}{
\caption{
Comparison of the pressure, energy and entropy density calculated with
HISQ with the HRG results.
}
\label{fig_eos}
}
\end{figure}

For 2+1 flavor QCD the equation of state is now available at
the physical pion mass in the continuum limit from the Wuppertal-Budapest
collaboration~\cite{Borsanyi:2013bia} that used stout fermions and the HotQCD
collaboration~\cite{Bazavov:2014pvz}
that used HISQ. A comparison of these results for the trace anomaly,
pressure and entropy density, $s=(\epsilon+p)/T$, is shown in Fig.~\ref{fig_e3p_p_s}.
In general, there is good agreement between them, with deviation of about $2\sigma$
in the integrated quantities arising around 400~MeV.
In Fig.~\ref{fig_eos} the HISQ results for the pressure,
energy and entropy density are compared with the HRG results. The yellow-shaded box
represents the chiral crossover region.

\begin{figure}
\centering
\includegraphics[width=0.47\textwidth]{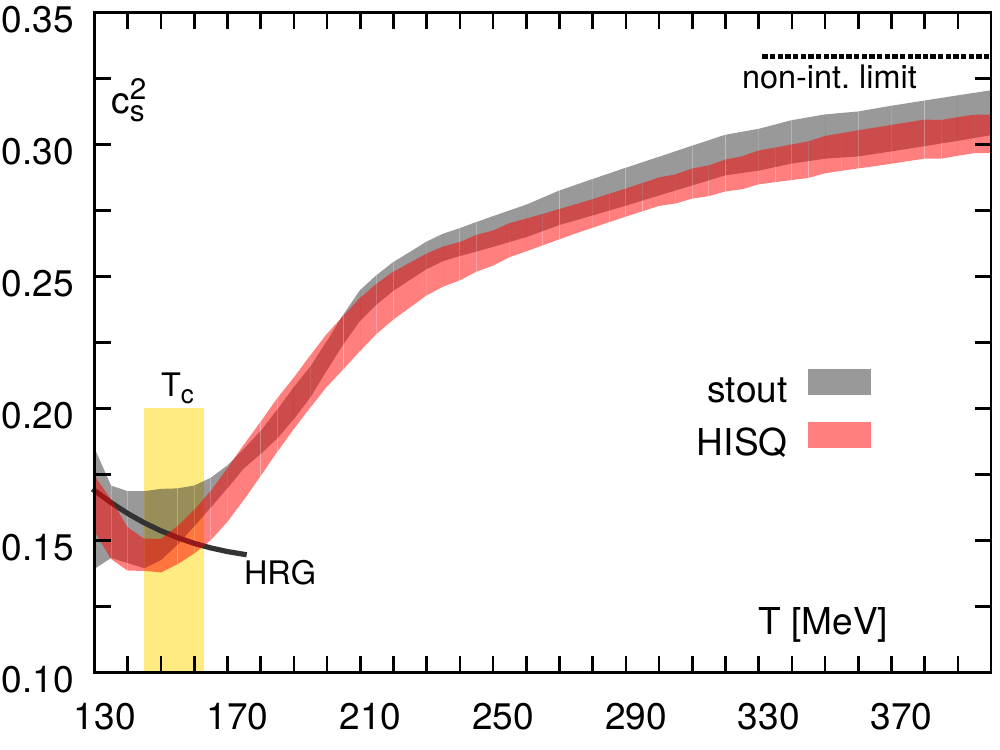}\hfill
\includegraphics[width=0.47\textwidth]{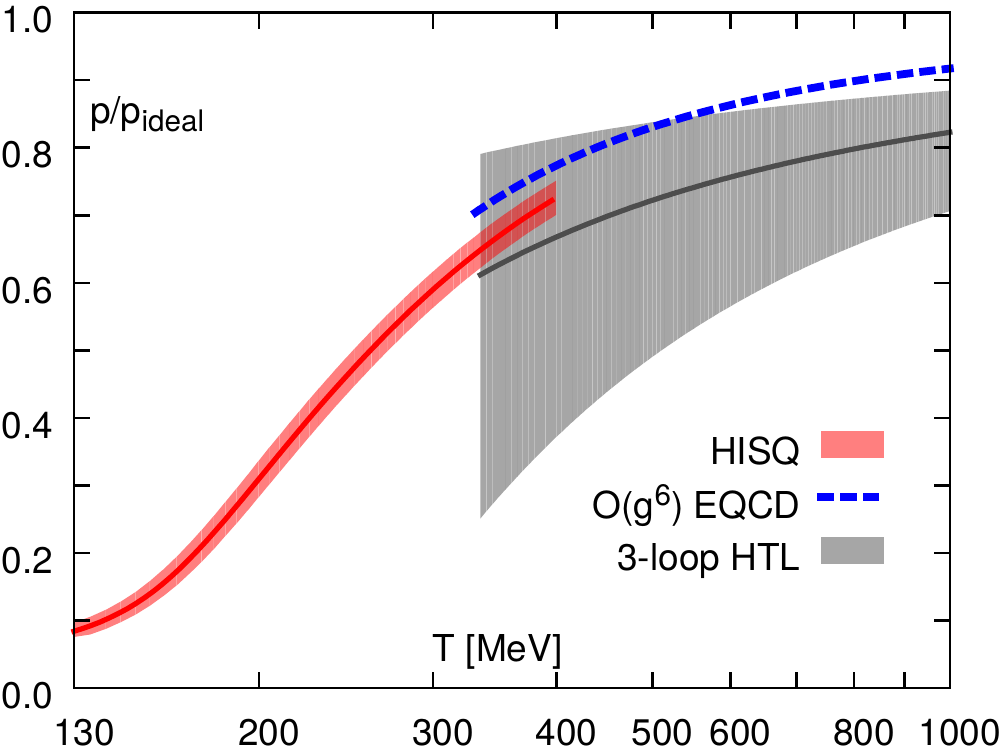}
\parbox[b]{0.46\textwidth}{
\caption{
The speed of sound for the stout and HISQ action compared with HRG.
}
\label{fig_cs2}
}
\hfill
\parbox[b]{0.46\textwidth}{
\caption{
Pressure calculated with HISQ, in the EQCD and
in HTL perturbation theory.
}
\label{fig_eos_pert}
}
\end{figure}

The second-order derivatives of the free energy can also be constructed,
the specific heat is
\begin{equation}
C_V=\left.\frac{\partial\epsilon}{\partial T}\right|_V
\equiv\left(4\frac{\epsilon}{T^4}
+\left.T\frac{\partial(\epsilon/T^4)}{\partial T}\right|_V
\right)T^3,
\end{equation}
where the second term is related to the trace anomaly and its derivative:
\begin{equation}
T\frac{d(\epsilon/T^4)}{dT}=3\frac{\Theta^{\mu\mu}}{T^4}
+T\frac{d(\Theta^{\mu\mu}/T^4)}{dT},
\end{equation}
and the speed of sound is
\begin{equation}
c_s^2=\frac{\partial p}{\partial\epsilon}=
\frac{\partial p/\partial T}{\partial\epsilon/\partial T}=\frac{s}{C_V}.
\end{equation}
The speed of sound is of phenomenological interest, because the softest point
of the equation of state, \textit{i.e.} where the speed of sound reaches the
minimum, corresponds to the temperature and energy density range where the system
spends longer time and the expansion and cooling of matter slows down.
For this reason one may expect to observe characteristic signatures from this
stage of the evolution of QGP in heavy-ion collisions~\cite{Hung:1994eq}.
The speed of sound is shown  in Fig.~\ref{fig_cs2}.
It appears that the softest point is reached
at the low side of the crossover region, at $T\simeq(145-150)$~MeV.

The lattice equation of state can now be compared with the perturbative
calculations to determine at what temperature weak-coupling expansions
may be trusted.
In Fig.~\ref{fig_eos_pert} the pressure is compared with the perturbative
calculations in the Hard Thermal Loop (HTL)~\cite{Haque:2014rua}
and Electrostatic QCD (EQCD)~\cite{Laine:2006cp} schemes.
Given the large uncertainty from varying the scale in the HTL calculation,
the results generally agree, but it
appears that lattice calculations at higher temperatures will be needed
to reliably connect the lattice equation of state to a perturbative one
and determine which, HTL or EQCD calculation, describes the equation of state
better.

\subsection{$N_f=2+1+1$}

\begin{figure}[b]
\centering
\includegraphics[width=0.47\textwidth]{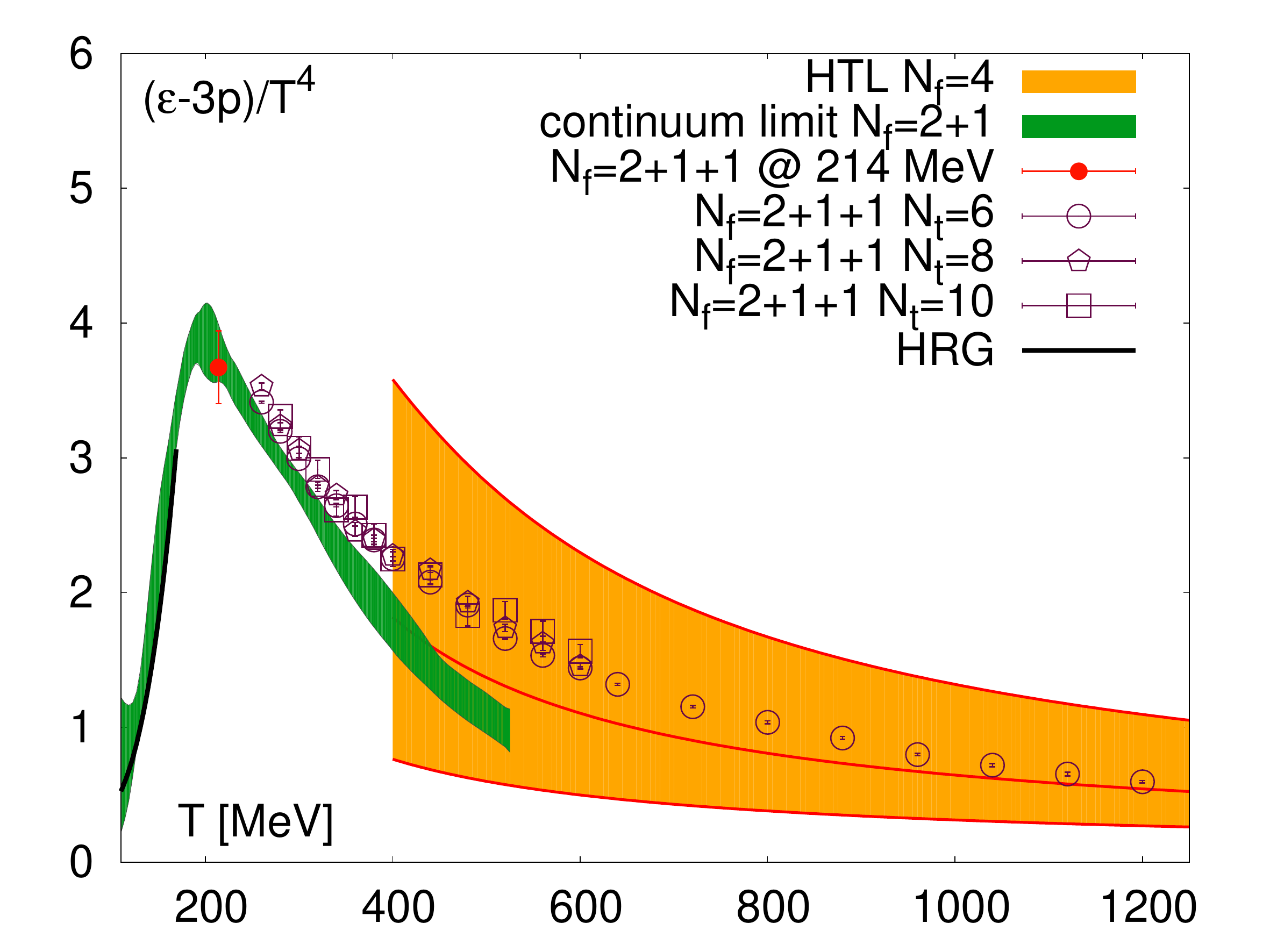}\hfill
\includegraphics[width=0.47\textwidth]{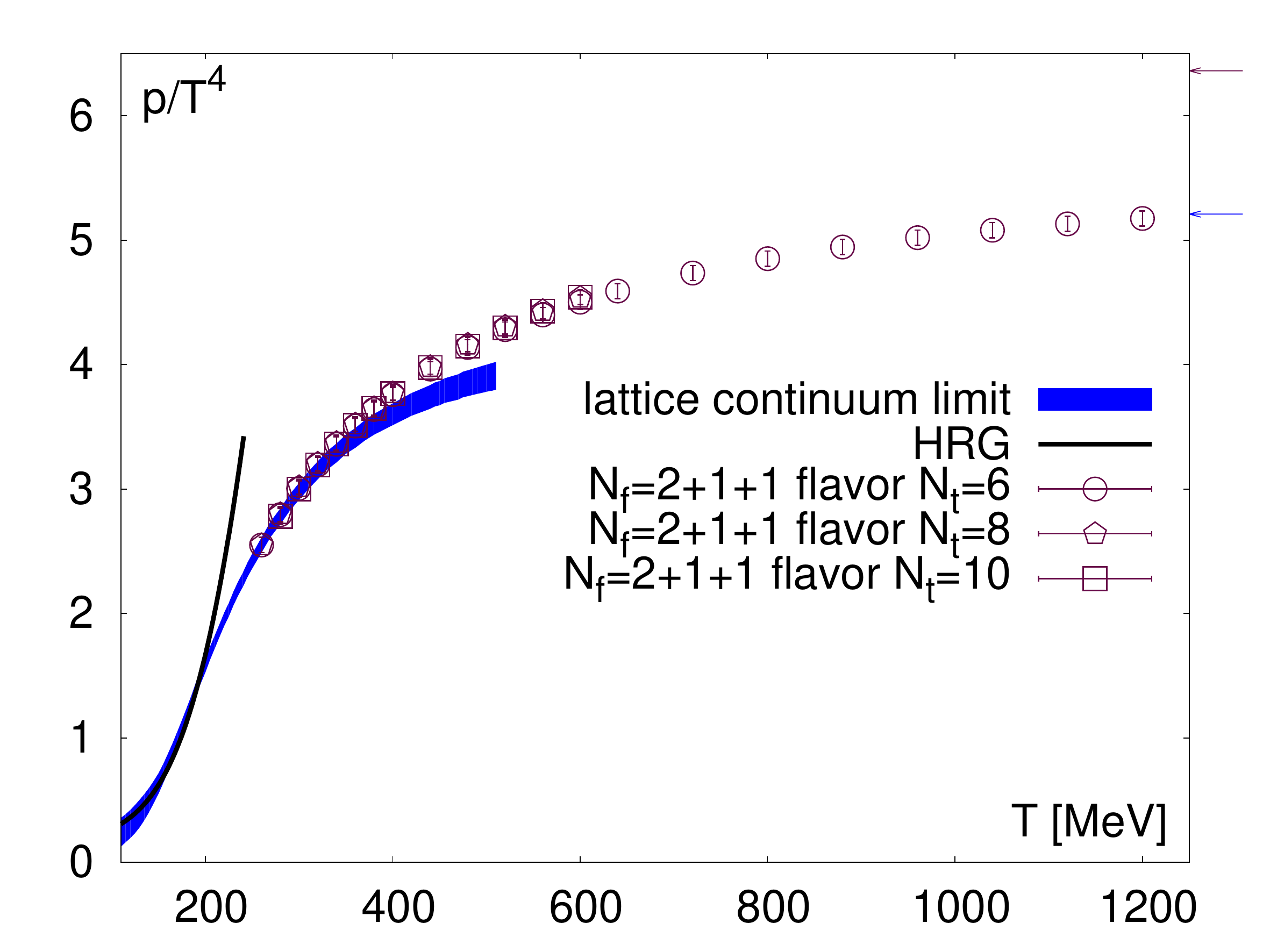}
\parbox[b]{0.46\textwidth}{
\caption{
The 2+1+1 flavor trace anomaly with the 4stout action together
with the 2+1 flavor continuum and HTL results.
}
\label{fig_eos_charm}
}
\hfill
\parbox[b]{0.46\textwidth}{
\caption{
Pressure in 2+1+1 flavor QCD with the 4stout action together 
with the 2+1 flavor continuum and HRG results.
}
\label{fig_p_charm}
}
\end{figure}

Apart from presenting their final result on the 2+1 flavor equation of
state~\cite{Borsanyi:2013bia}, the Wuppertal-Budapest collaboration
has also extended their calculation of the 2+1+1 flavor equation
of state~\cite{Borsanyi:2014rza}. For the latter case they use the stout
action with four levels of smearing, called 4stout (while two levels of smearing were
used in the previous work). They show that at $T=214$~MeV the value
of the trace anomaly with and without the dynamical charm is the same,
in line with the perturbative
observation that the charm contribution starts to matter around
300~MeV~\cite{Laine:2006cp}. The 2+1+1 flavor trace anomaly with the
4stout action at $N_\tau=6$, $8$ and $10$ is shown in Fig.~\ref{fig_eos_charm}
and compared with the continuum 2+1 flavor result and the perturbative
HTL calculation. To cover a wide temperature range $300<T<1000$~MeV
and set the line of constant physics, the range of gauge couplings is split into
three regions. The lattice scale is then set in different ways: from spectroscopy at
coarse lattices, from simulations at the $m_l=m_s$ flavor symmetric point at
intermediate lattices and from the $w_0$ scale determined
with the Wilson flow~\cite{Luscher:2010iy} at fine lattices.
The results for the pressure are shown in Fig.~\ref{fig_p_charm}.

\subsection{$N_f=2+1$, $\mu_B>0$}

\begin{figure}
\centering
\includegraphics[width=0.47\textwidth]{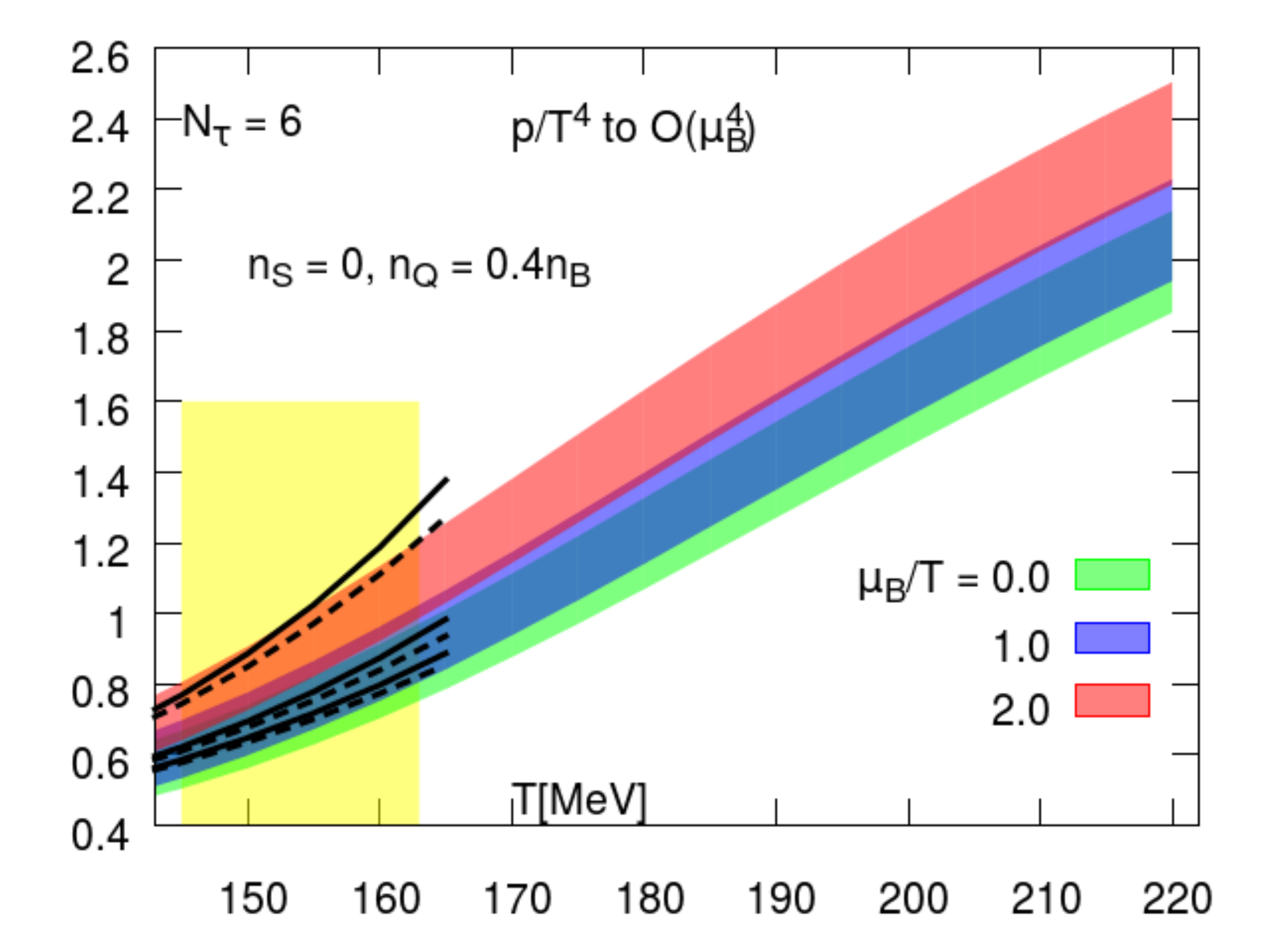}\hfill
\includegraphics[width=0.47\textwidth]{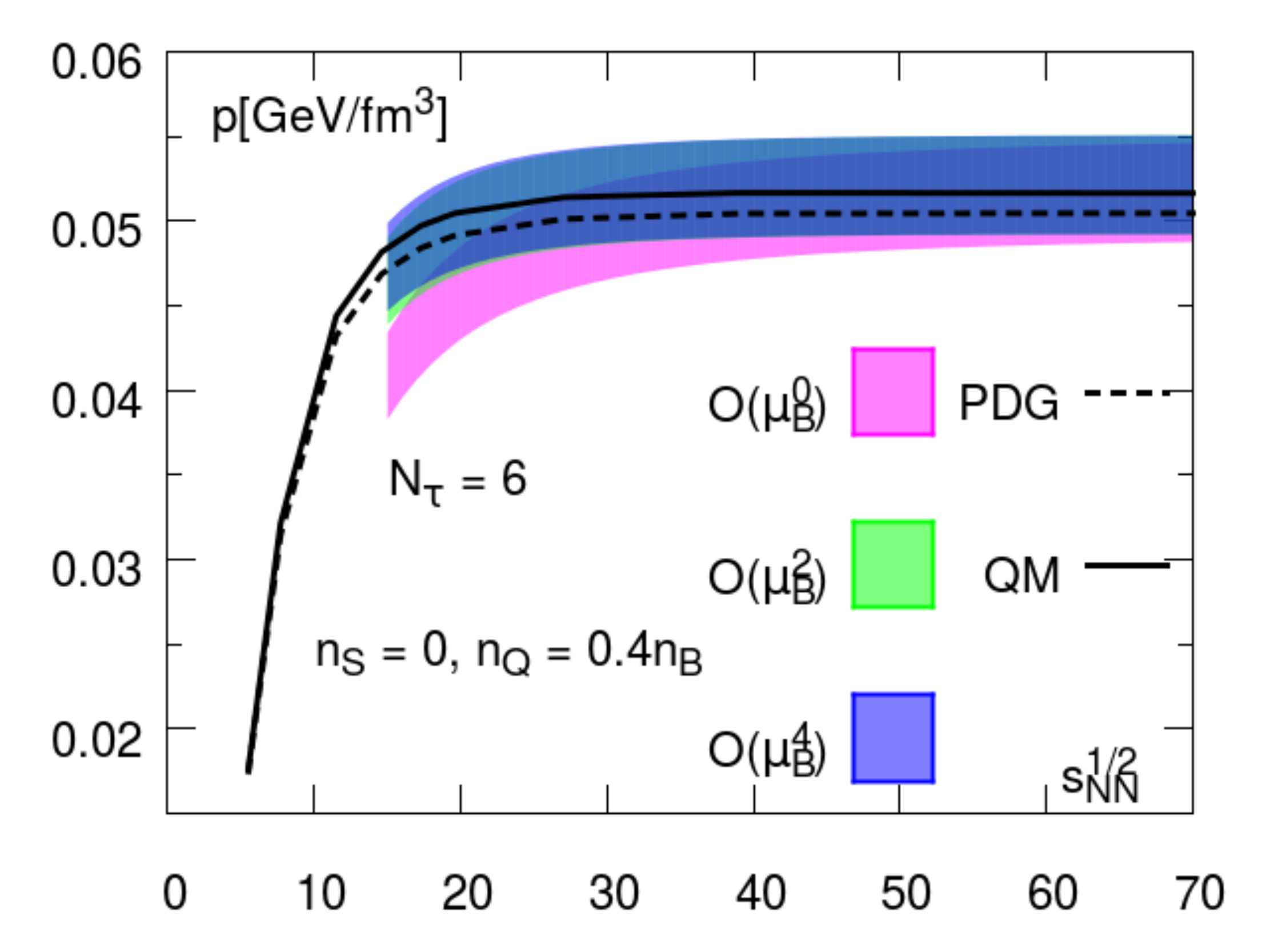}
\caption{
Left: Pressure at $O(\mu_B^4)$ on $N_\tau=6$ lattice with HISQ. The electric
charge to baryon number ratio is set to 0.4, similar to the one encountered in
heavy-ion collision experiments. Right: Pressure at various orders in $\mu_B$
on the freeze-out curve, on $N_\tau=6$ lattice with HISQ.
}
\label{fig_eos_mu4}
\end{figure}

The susceptibilities defined in Eq.~(\ref{eq_gen}) are the coefficients
of the Taylor expansion of the pressure with respect to the chemical
potentials, and they can be used to extend the equation of state to
non-zero baryon chemical potential. The BNL-Bielefeld-CCNU collaboration
calculated the pressure and energy density at $O(\mu_B^4)$
at several values of the chemical potential up to $\mu_B/T=2$,
by measuring the second and fourth order susceptibilities to
very high precision at $N_\tau=6$ and $8$ lattices with
the HISQ action~\cite{Hegde:2014wga}.
The results for the pressure are shown in Fig.~\ref{fig_eos_mu4} (left).
By using a parametrization of the freeze-out curve that relates
$(T^f,\mu_B^f)$ to the beam energy $s_{NN}^{1/2}$~\cite{Cleymans:1999st}
the equation of state
can also be calculated along that curve. The pressure on the freeze-out
curve is shown in Fig.~\ref{fig_eos_mu4} (right). The fourth-order
equation of state can be useful down to beam energy $s_{NN}^{1/2}\sim20$~GeV,
while below that energy higher-order terms will be needed for correct
description.

\section{In-medium properties of mesons}

While the dynamics of the chiral crossover in QCD is determined by
the light degrees of freedom, heavy quarks play a special role
in understanding the properties of the deconfined medium.
In heavy-ion collisions heavy quarks are created at the early stages
of the quark-gluon plasma formation and can serve as probes
of the medium. Melting of the heavy-quark bound states due to the
color screening in the plasma was suggested as a signature of
QGP formation~\cite{Matsui:1986dk}.

Consider a local mesonic operator of the form:
\begin{equation}
J_H(t,\vec{x})=\bar q(t,\vec{x}) \Gamma_H q(t,\vec{x})
\label{eq_meson}
\end{equation}
where $q(t,\vec{x})$ is the quark field operator and $\Gamma_H$
denotes possible $\gamma$-matrix structure of the state
(\textit{e.g.} pseudoscalar, vector, etc.).

Defining the real time two-point functions of the
currents~(\ref{eq_meson}):
\begin{equation}
D^{>}_H(t,\vec{x}) = \langle
J_H(t, \vec{x}) J_H(0, \vec{0})\rangle,\,\,\,\,\,
D^{<}_H(t,\vec{x}) =
\langle J_H(0, \vec{0}) J_H(t,\vec{x}) \rangle , t>0
\end{equation}
and making the Fourier transform
\begin{equation}
 D^{>(<)}_H(\omega, \vec{p}) = \int_{-\infty}^{\infty} dt \int d^3 x 
 e^{i \omega t -i \vec{p} \cdot \vec{x}} D^{>(<)}_H(t,\vec{x})
\label{eq_ft}
\end{equation}
we can define the spectral function
\begin{equation}
\rho(\omega,\vec{p}) = \frac{1}{2 \pi} (D^{>}_H(\omega, \vec{p})-D^{<}_H(\omega, \vec{p})).
\end{equation}
Most of the dynamic properties of the state and its behavior in the medium
are encoded in the spectral function. A stable state of mass $M$ contributes
a $\delta$-function peak of the form:
\begin{equation}
\rho(\omega,\vec{p}) = | \langle 0 | J_H | H \rangle |^2 \theta(\omega)
\delta(p^2 - M^2).
\label{eq_stable}
\end{equation}
Broadening of such a peak with increasing temperature describes thermal
modification of the state, and its disappearance signals melting
of the state. Evaluation of the spectral function requires calculating
real time two-point functions, to which lattice does not have direct access.
Instead, lattice simulations are done in Euclidean space-time and produce
Euclidean correlation functions:
\begin{equation}
G(\tau, \vec{p}) = \int d^3x e^{i \vec{p}.\vec{x}} 
\langle J_H(\tau, \vec{x}) J_H(0,
\vec{0}) \rangle,
\end{equation}
which are analytic continuations of $D^{>}_H(t,\vec{p})$ as
\begin{equation}
G(\tau,\vec{p})=D^{>}_H(-i\tau,\vec{p}).
\label{cont}
\end{equation}
The relation between the Euclidean correlator and the spectral function
is more complicated:
\begin{eqnarray}
G(\tau, \vec{p}) &=& \int_0^{\infty} d \omega
\rho(\omega,\vec{p}) K(\omega, \tau), \label{eq.spect} \label{eq_corr}\\
K(\omega, \tau) &=& \frac{\cosh(\omega(\tau-1/2
T))}{\sinh(\omega/2 T)},
\label{eq_kernel}
\end{eqnarray}
and, unlike the Fourier transform, does not allow for direct inversion.

To reconstruct the spectral function $\rho(\omega,\vec{p})$ from the correlator in
the l.h.s. of Eq.~(\ref{eq_corr}) (or, in other words, solve the integral
equation for $\rho$), Bayesian techniques are often employed, that
try to maximize the conditional probability $P[\rho|DH]$ that $\rho(\omega,\vec{p})$ is
the correct spectral function given data $D$ and some prior knowledge $H$.
One of such techniques, often applied to this problem,
is the Maximum Entropy Method (MEM)~\cite{Asakawa:2000tr}.
In this method the Shannon-Janes entropy:
\begin{equation}
S=\int d\omega\left[\rho(\omega)-m(\omega)-\rho(\omega)
\ln\left(\frac{\rho(\omega)}{m(\omega)}\right)\right]
\end{equation}
incorporates the positivity of the spectral function and all other prior
knowledge is parametrized by $m(\omega)$, which is called the default model.
The conditional probability is then
\begin{equation}
P[\rho|DH]=\exp\left(-\frac{1}{2}\chi^2+\alpha S\right),
\end{equation}
where $\alpha$ is a real parameter.

\subsection{Charmonium spectral functions}

\begin{figure}
\centering
\includegraphics[width=0.43\textwidth]{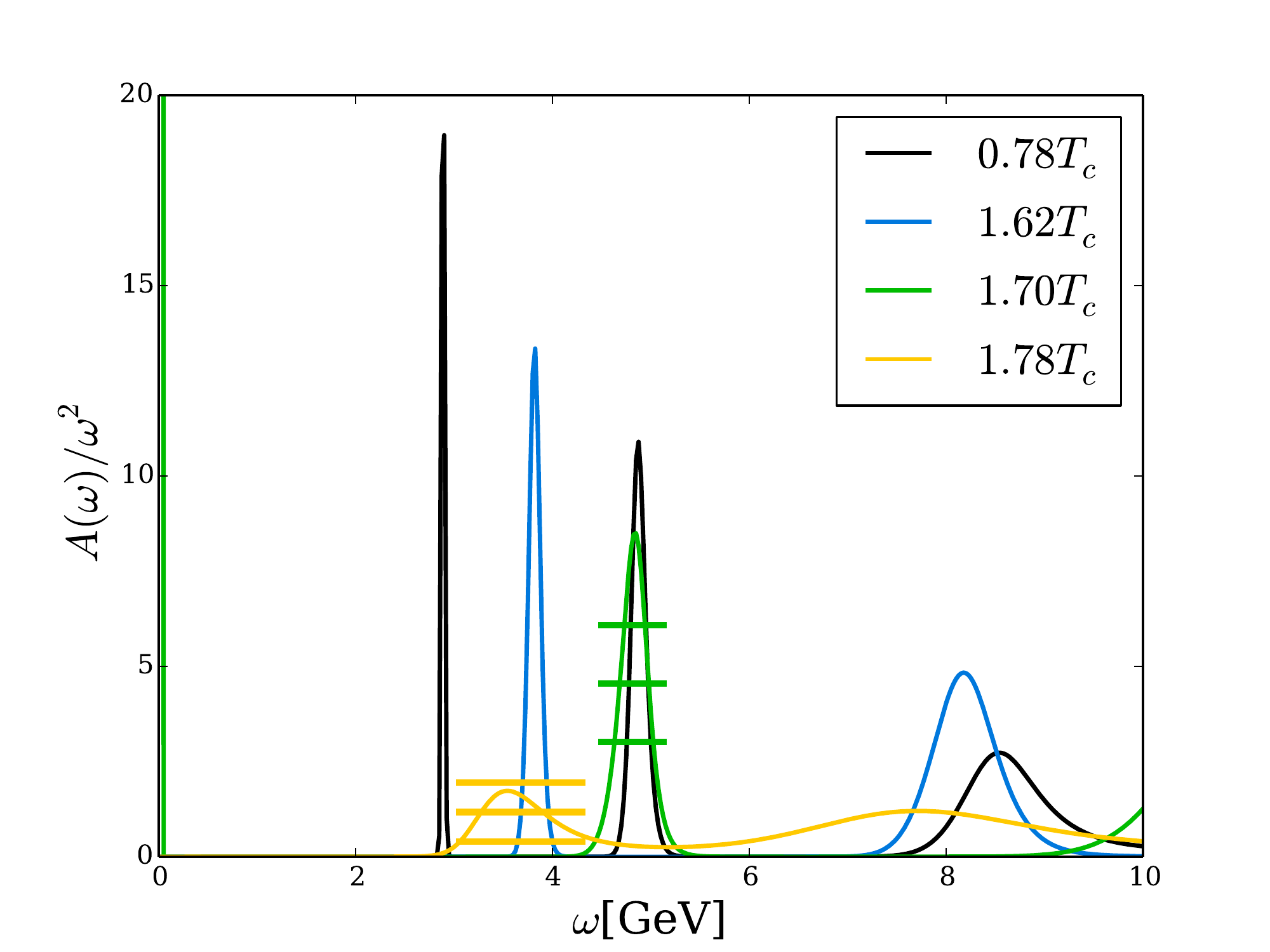}\hfill
\includegraphics[width=0.51\textwidth]{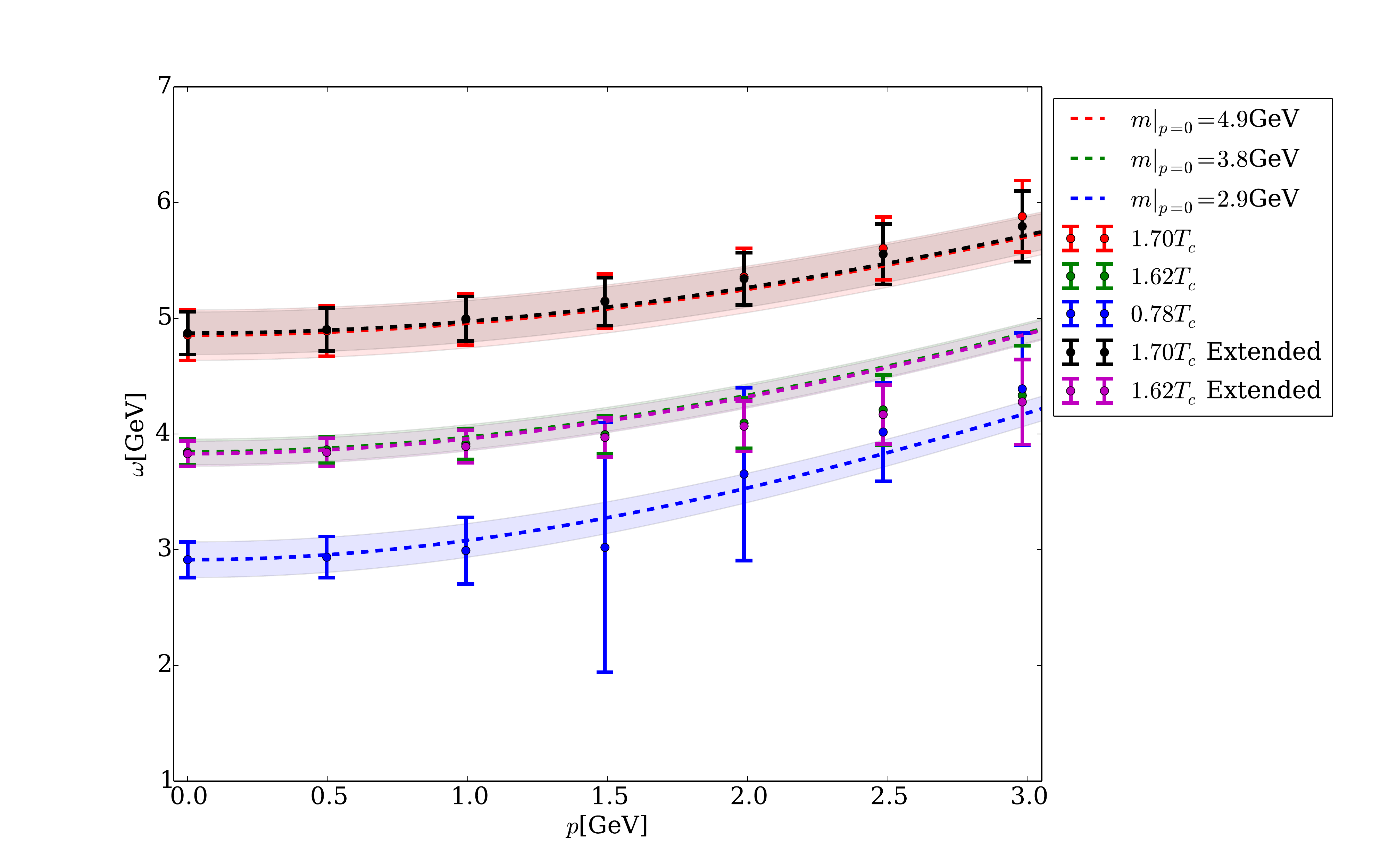}
\caption{
Left: Spectral function in the $\eta_c$ channel for several values of $T$.
Right: Dispersion relation of $\eta_c$ at three temperatures~\cite{Ikeda:2014vca}.
}
\label{fig_sp_charm1}
\end{figure}

Ikeda et al.~\cite{Ikeda:2014vca} developed a modification of MEM to study charmonium
spectral functions at finite momentum. Using quenched QCD configurations
and Wilson fermions in the valence sector they calculated charmonium
correlators in the pseudoscalar channel ($\eta_c$). Working on anisotropic
lattices $a_\sigma/4=a_\tau=0.00975$~fm they covered a temperature range
$(0.78-1.78)T_c$. The spectral function of $\eta_c$ at several temperatures
is shown in Fig.~\ref{fig_sp_charm1} (left). Presence of a well pronounced
peak at $1.7T_c$ is interpreted as survival of the $\eta_c$ state up to
that temperature. Having the spectral function allows for calculating
the dispersion relation, defined in this case as the dependence of the
peak position on momentum $p$. This dispersion relation is shown in
Fig.~\ref{fig_sp_charm1} (right) and it seems to be described by
the vacuum form $\omega=\sqrt{m^2+p^2}$ up to $1.7T_c$.

\begin{figure}
\centering
\includegraphics[width=0.48\textwidth]{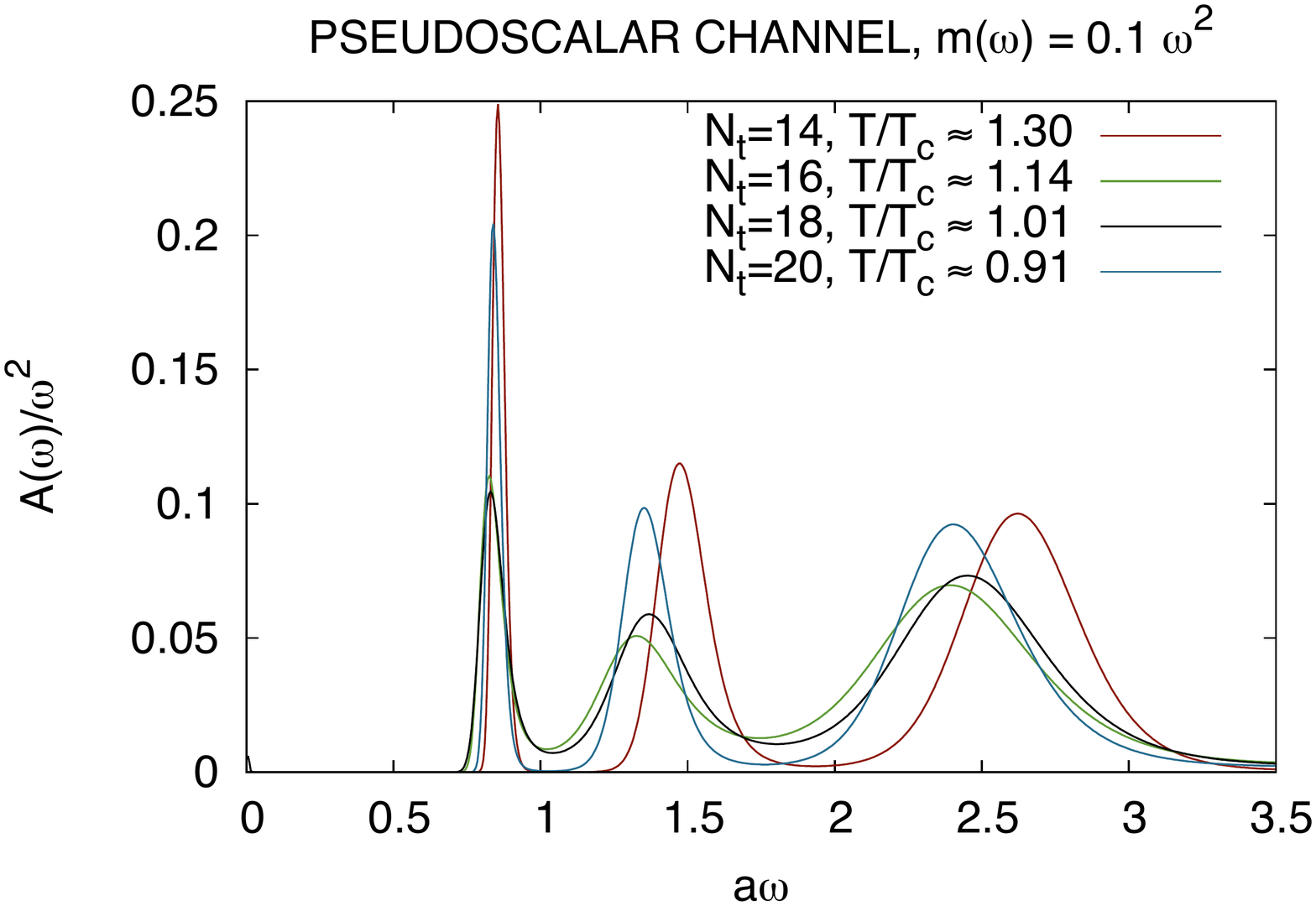}\hfill
\includegraphics[width=0.48\textwidth]{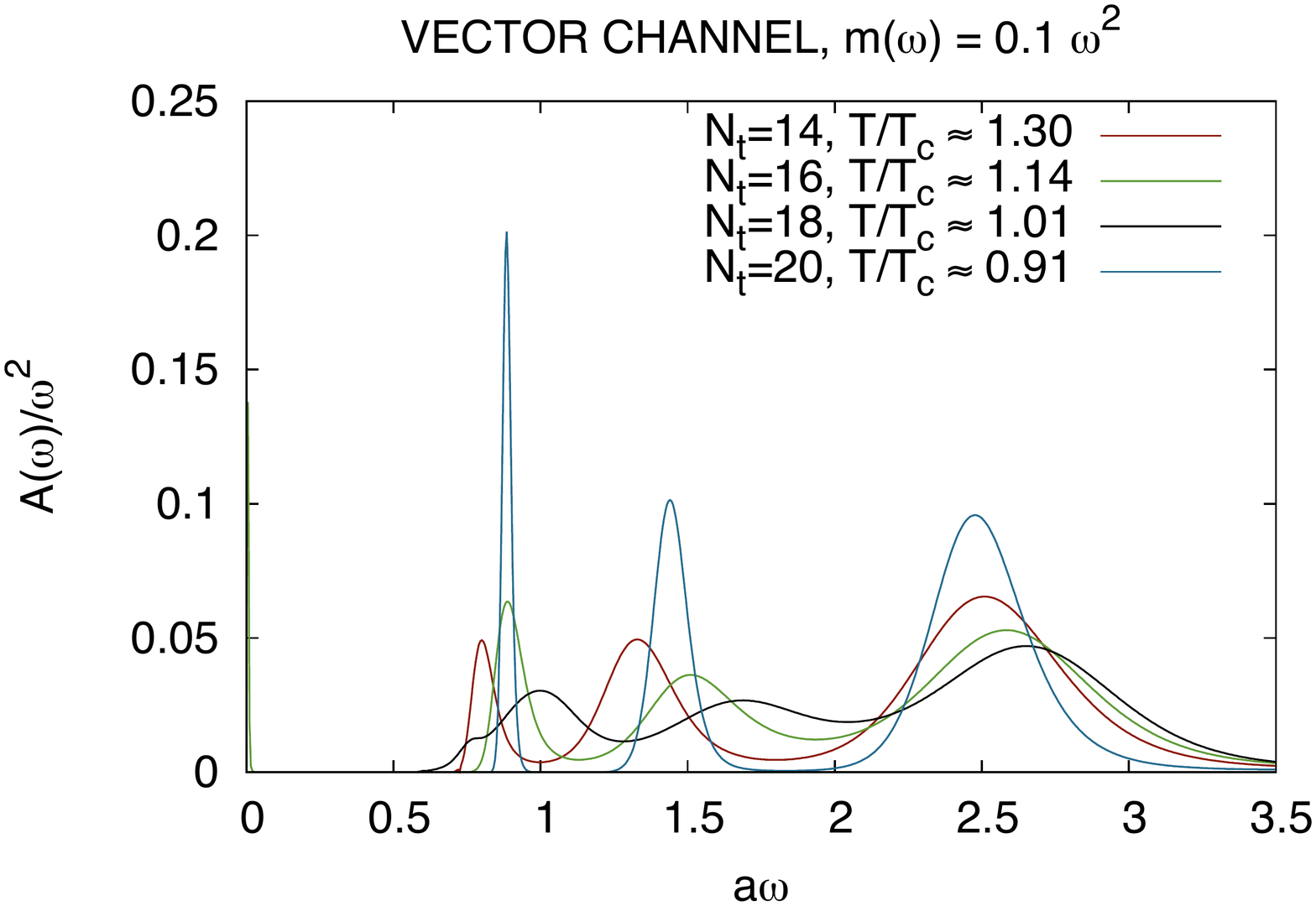}
\caption{
Temperature dependence of spectral functions in the $\eta_c$ (left)
and $J/\psi$ (right) channel~\cite{Borsanyi:2014pta}.
}
\label{fig_sp_charm2}
\end{figure}

The Wuppertal-Budapest collaboration calculated the charmonium spectral
functions in the pseudoscalar and vector channels using 2+1 flavors of
dynamical Wilson quarks with the pion mass $m_\pi=545$~MeV~\cite{Borsanyi:2014pta}.
They employed
isotropic lattices down to $0.057$~fm. The results are shown in
Fig.~\ref{fig_sp_charm2}. Apart from the reconstruction of the spectral
functions with MEM, the analysis of the Euclidean correlators was also
performed, leading to a conclusion that no melting of $\eta_c$ and $J/\psi$
mesons was observed up to a temperature of $1.4T_c$.

\subsection{Bottomonium spectral functions}

The bottom quark is substantially heavier and controlling the discretization
errors is harder, therefore effective theories are often used.

\begin{figure}
\centering
\includegraphics[width=0.472\textwidth]{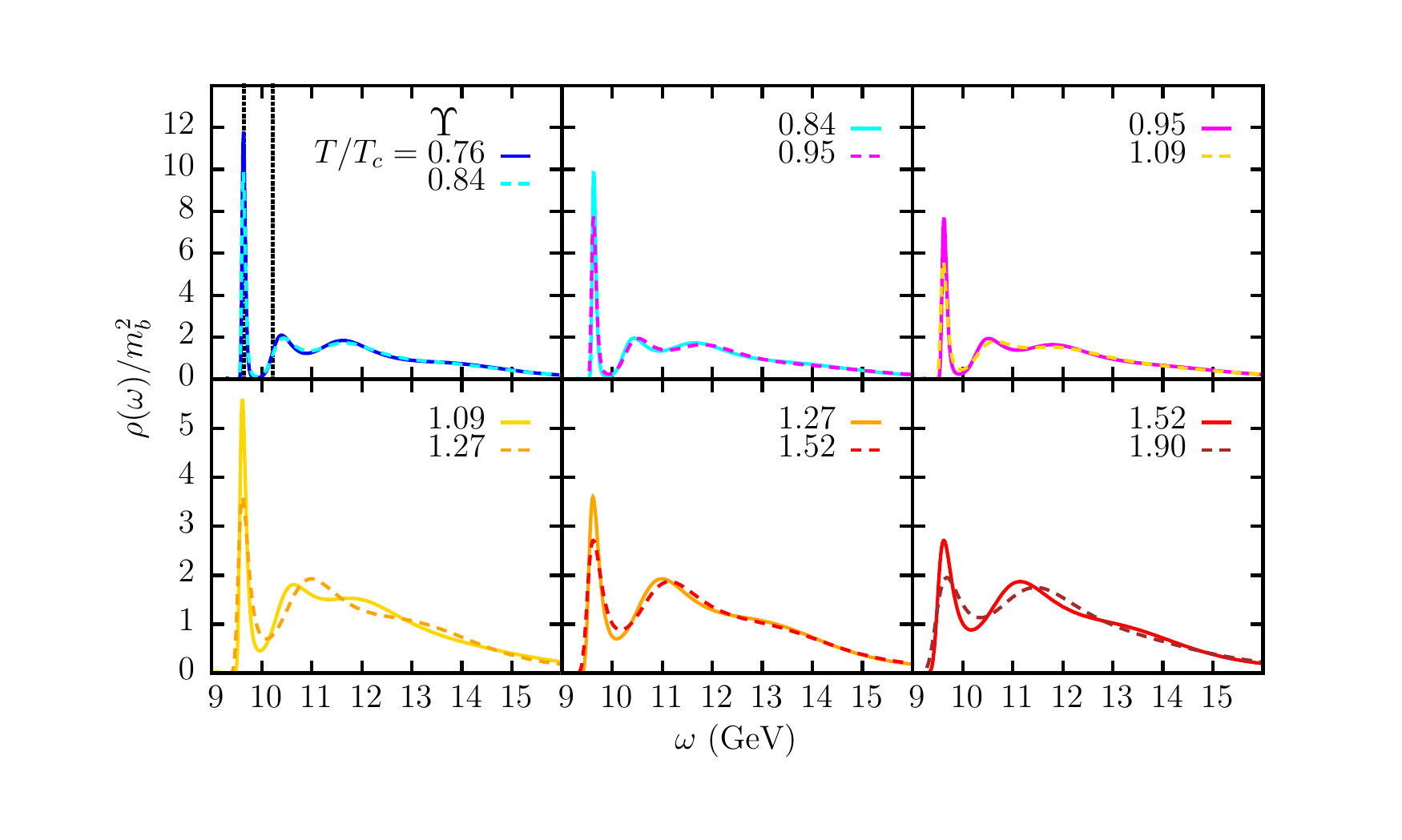}\hfill
\includegraphics[width=0.50\textwidth]{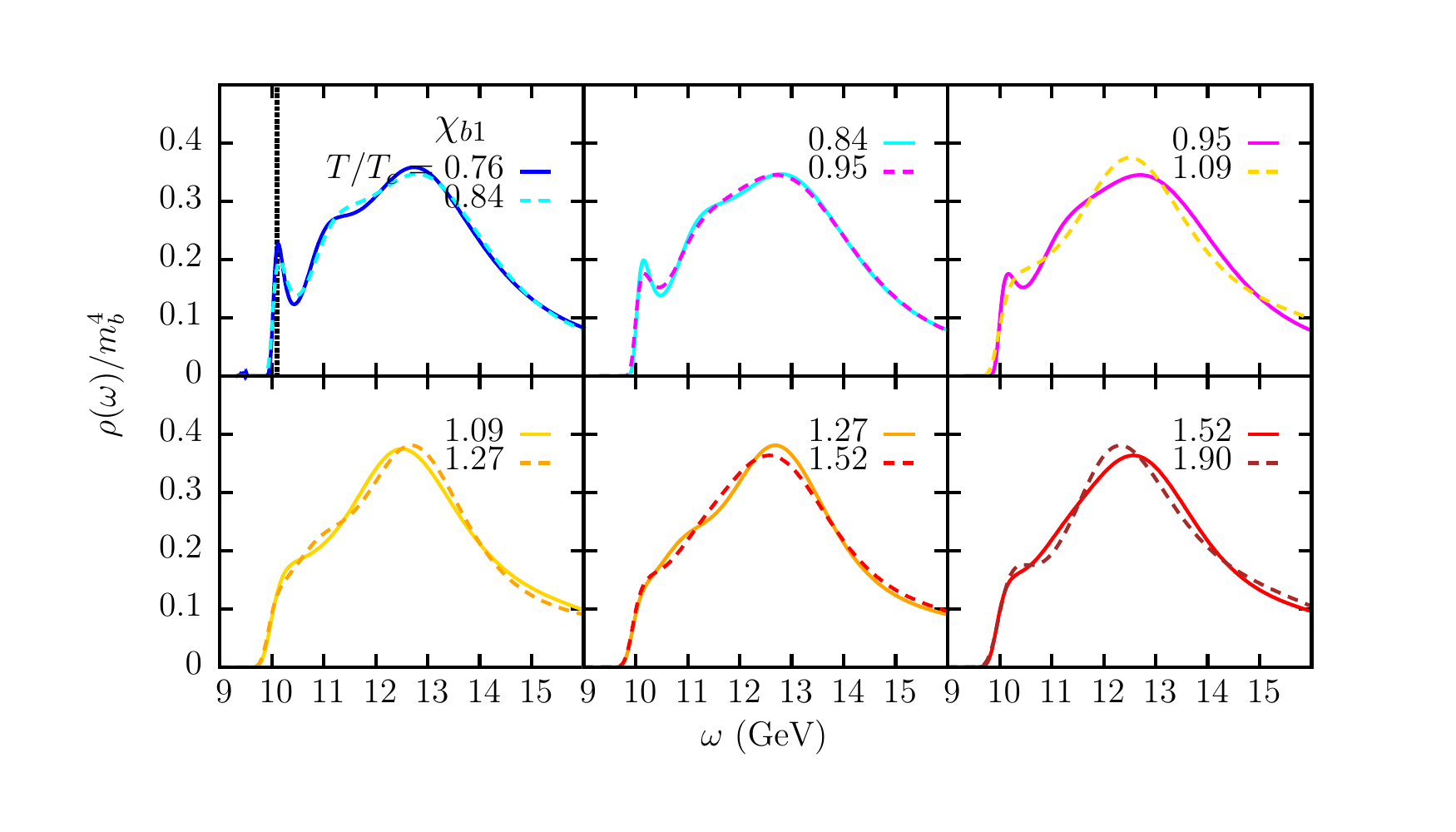}
\caption{
Temperature dependence of spectral functions in the $\Upsilon$ (left)
and $\chi_{b1}$ (right) channel~\cite{Aarts:2014cda}.
}
\label{fig_sp_bottom1}
\end{figure}

The FASTSUM collaboration performed a calculation of the bottomonium spectral
functions with MEM using Non-Relativistic QCD (NRQCD)~\cite{Aarts:2014cda} for the bottom
quark. For the gauge background they used 2+1 flavors of clover improved
Wilson quarks with the pion mass $m_\pi=400$~MeV. The spectral functions
for $S$- and $P$-wave channels ($\Upsilon$ and $\chi_{b1}$ mesons) are shown
in Fig.~\ref{fig_sp_bottom1}. The ground state peak in the $S$-wave case, present
at all temperatures, indicates the survival of $\Upsilon$ up to the highest
temperature of $1.9T_c$. In the $P$-wave case disappearance of the
peak is observed slightly above $T_c$, indicating dissociation of $\chi_{b1}$
once the deconfined phase is reached.

\begin{figure}
\centering
\includegraphics[height=0.47\textwidth,angle=-90]{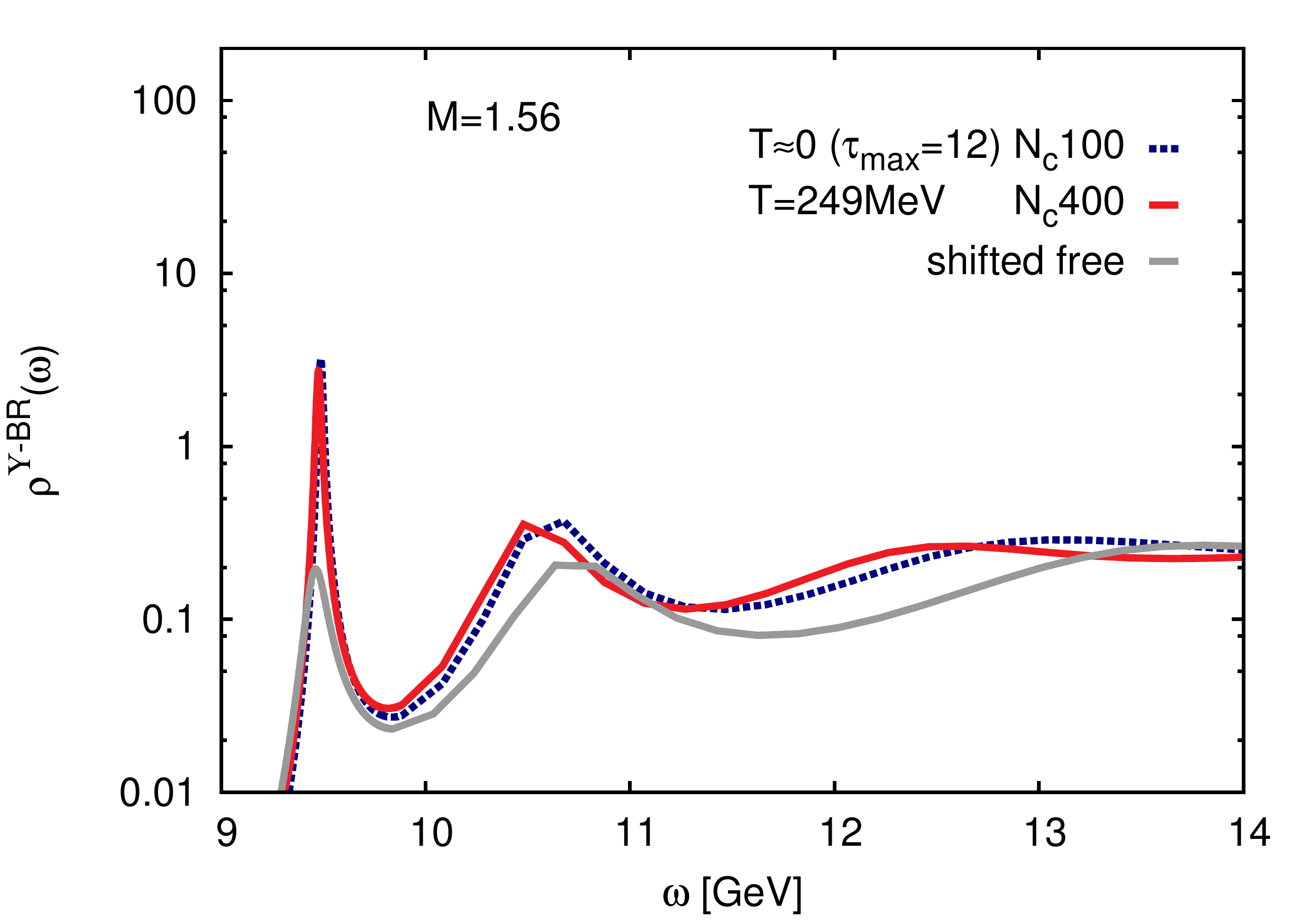}\hfill
\includegraphics[height=0.47\textwidth,angle=-90]{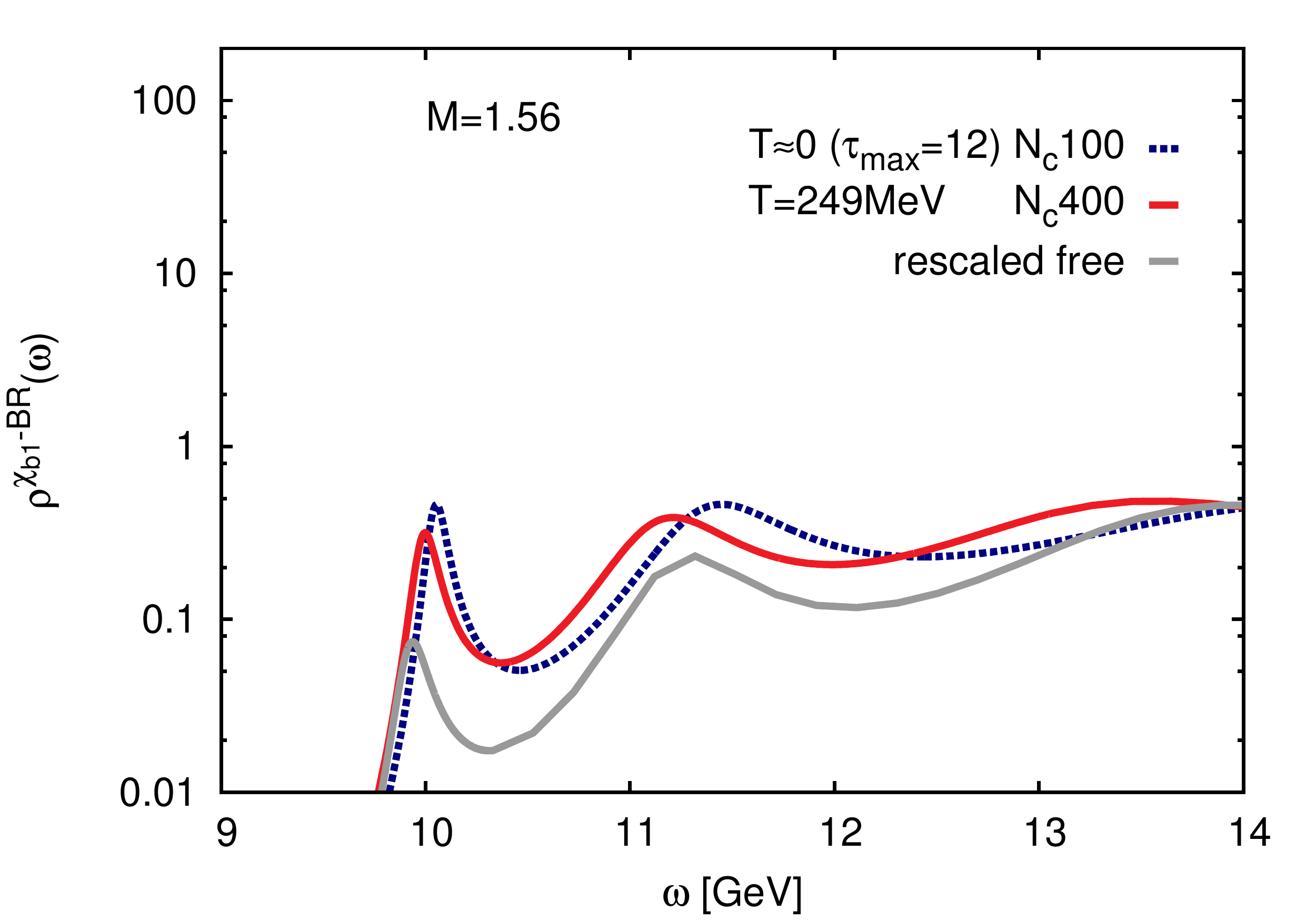}
\caption{
$\Upsilon$ (left) and $\chi_{b1}$ (right) spectral functions
at $T=0$ and $249$~MeV~\cite{Kim:2014iga,Kim:2014nda}.
}
\label{fig_sp_bottom2}
\end{figure}

Kim et al.~\cite{Kim:2014iga,Kim:2014nda} calculated the bottomonium spectral functions
also employing NRQCD for the bottom quark, but on a different gauge background of
2+1 flavors of highly improved staggered quarks with the pion mass $m_\pi=160$~MeV.
They also used a different Bayesian approach, recently suggested in Ref.~\cite{Burnier:2013nla}.
The $S$- and $P$-wave spectral functions at $T=0$ and $T=249$~MeV are shown
in Fig.~\ref{fig_sp_bottom2}. In this case, contrary to the finding of the
previous group, the $P$-wave state features a peak at $T=1.6T_c$ (taking
$T_c=154$~MeV), signaling no dissociation of $\chi_{b1}$ in the plasma up
to that temperature. Clearly, further studies are needed to resolve
the tension between these two results.

\subsection{Static quark potential at finite temperature}

\begin{figure}
\centering
\includegraphics[width=0.96\textwidth]{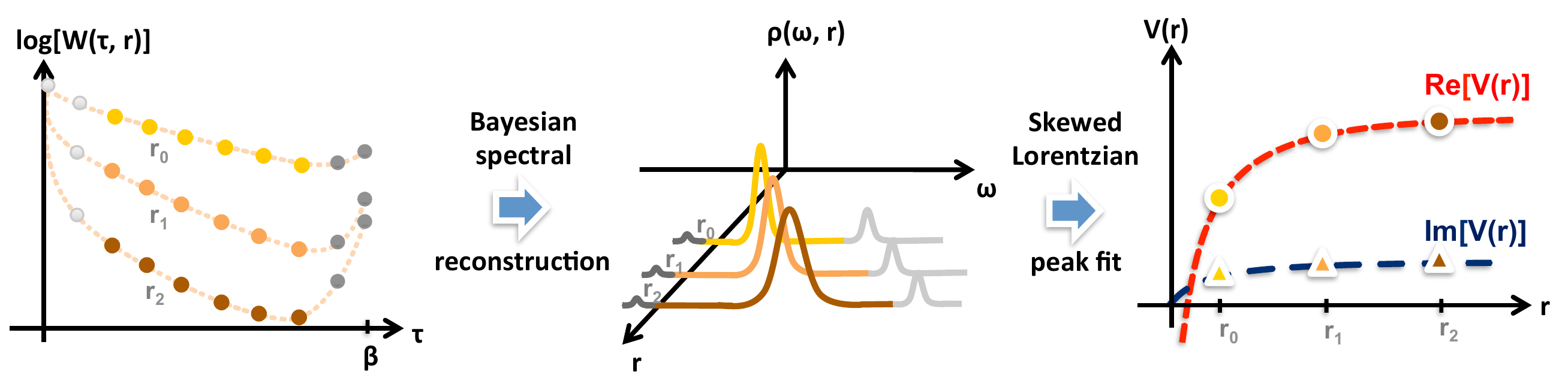}
\caption{
The strategy for extracting the imaginary part of the static quark potential
by reconstructing the spectral function of a Wilson loop, from~\cite{Burnier:2014yda}.
}
\label{fig_strategy}
\end{figure}

\begin{figure}[b]
\centering
\includegraphics[width=0.47\textwidth]{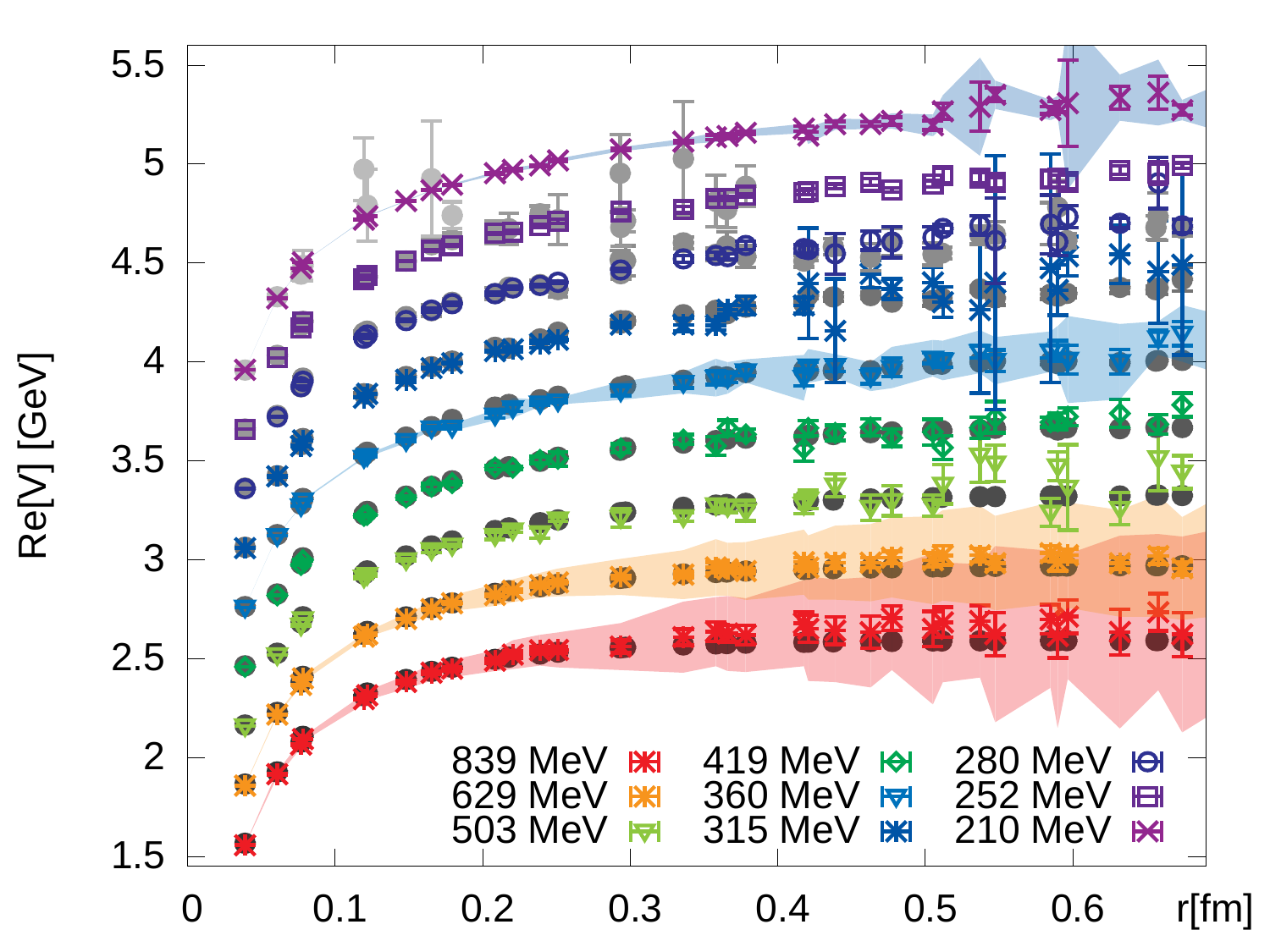}\hfill
\includegraphics[width=0.47\textwidth]{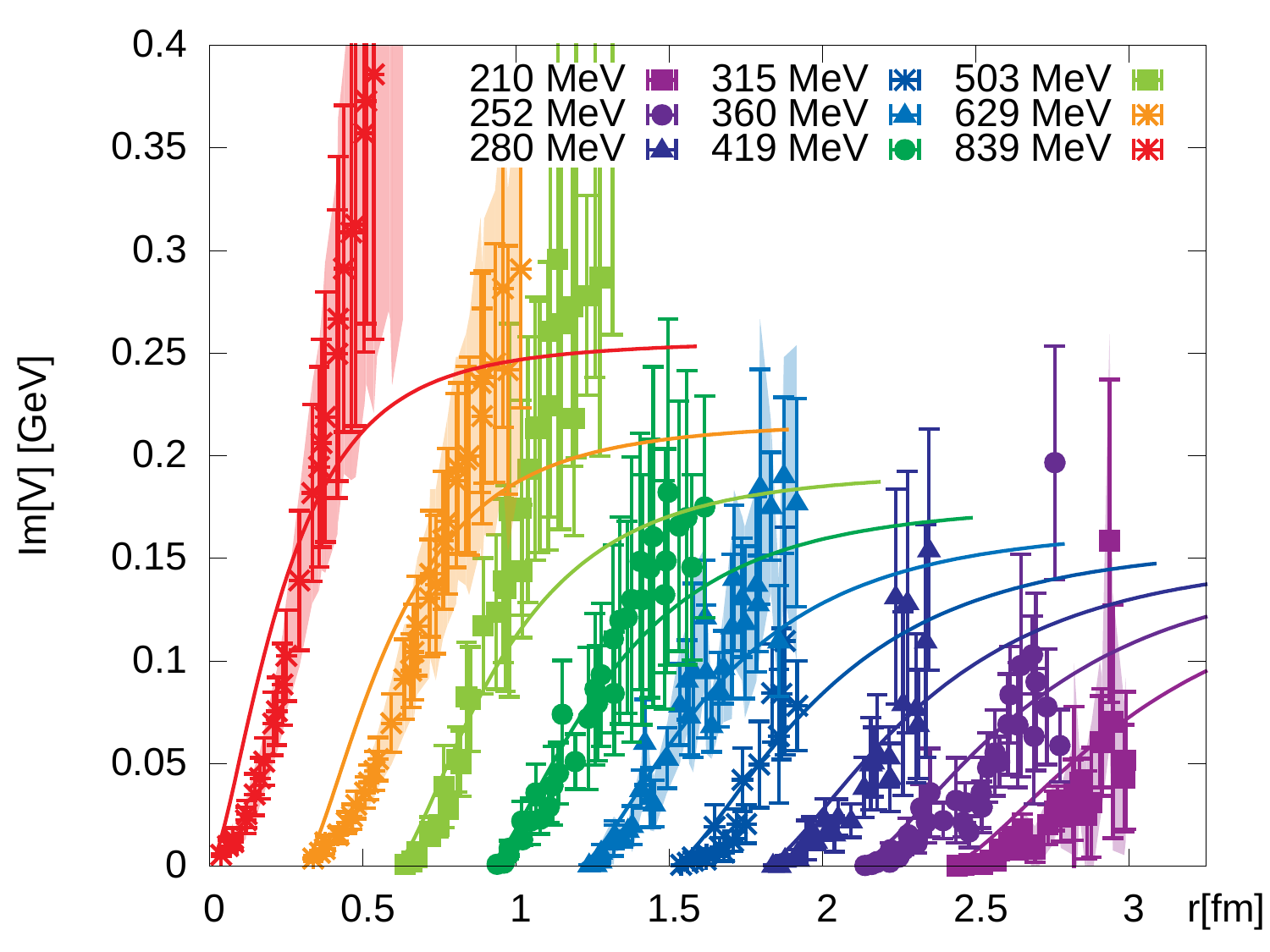}
\caption{
Real (left) and imaginary (right) part of the static quark potential
in $SU(3)$ pure gauge theory.
}
\label{fig_re_im}
\end{figure}

The static quark potential at zero temperature is a well-known quantity
that has been studied since the early days of lattice QCD. It is often
used to set the scale in lattice calculations, since its measurement
is numerically cheap. At finite temperature the color singlet free energy
of a static quark anti quark pair is sometimes used as a proxy for it.
The situation is, however, more involved. The potential is related
to the late real-time behavior of the rectangular Wilson loop:
\begin{equation}
V(r)=\lim_{t\to\infty} \frac{i\partial_t W(t,r)}{W(t,r)}\label{Eq:VRealTimeDef}, \quad W(t,r) = \left\langle {\rm exp}\Big[ - {ig}\int_{\square} dx_\mu A^\mu(x) \Big] \right\rangle,
\end{equation}
and, in general, it is complex valued. At finite temperature, the familiar real
part describes the Debye color screening, while the imaginary part is related
to the Landau damping in the plasma.
Since the real-time behavior is not accessible in Euclidean lattice formalism,
the imaginary part of the potential cannot be measured directly in Monte Carlo
simulations. However, a strategy similar to the one used for spectral functions
can be applied, as sketched in Fig.~\ref{fig_strategy}.
By finding the spectral decomposition of the Euclidean Wilson loop
\begin{equation}
W(\tau,r)=\int d\omega e^{-\omega \tau} \rho(\omega,r)
\end{equation}
one can use the spectral function to calculate the complex-valued potential
\begin{equation}
V(r)=\lim_{t\to\infty}\frac{\int d\omega\, \omega e^{-i\omega t} \rho(\omega,r)}
{\int d\omega\, e^{-i\omega t} \rho(\omega,r)}.
\end{equation}
Burnier et al.~\cite{Burnier:2014yda,Burnier:2014ssa} calculated Wilson line correlators
in the Coulomb gauge (which are less noisy than Wilson loops) on quenched and
dynamical 2+1 flavor asqtad lattices, covering the temperature range
$210\leqslant T\leqslant839$~MeV and $148\leqslant T\leqslant 248$~MeV,
respectively. The results for the real and imaginary part of the potential
for the quenched case are shown in Fig.~\ref{fig_re_im}.
The real part is compared with the color singlet free energy, calculated on
the lattice, and the imaginary part with the HTL result at leading order.
The results for the imaginary part are consistent with the ones obtained recently
by using HTL-inspired spectral functions and fitting their parameters to measured
Wilson line correlators~\cite{Bazavov:2014kva}.

\section{Summary}

There have been many interesting developments in finite-temperature lattice
QCD in 2014, and some of them, summarized in this review, are listed below.
Calculations with domain-wall fermions confirm the staggered result on the chiral crossover
temperature $T_c$ and favor restoration of the axial symmetry considerably above $T_c$.
Fluctuations and correlations of conserved charges can be used to construct observables
that relate to the ones measured in heavy-ion collision experiments and, in particular,
efforts to calculate higher order cumulants of the electric charge are ongoing.
The results on the equation of state in 2+1 flavor QCD at the physical pion mass
in the continuum limit are now available and agreement between the stout and HISQ
calculations is demonstrated in $130-400$~MeV range. A new method for calculating
the equation of state on the lattice using non-conventional
boundary conditions has been introduced and tested for $SU(3)$ pure gauge theory.
Several groups continue calculations of the spectral functions for heavy quarkonia
and new techniques for extracting spectral functions have been introduced.

\acknowledgments

I am grateful to Chris Allton, Szabolcs Borsanyi, Bastian Brandt, Masanori Hanada, Tim Harris, Tamas Kovacs, Richard Lau, Florian Meyer, Michael Mueller-Preussker, Yoshifumi Nakamura, Daniel Nogradi, Marco Panero, Michele Pepe, Giancarlo Rossi, Alexander Rothkopf, Finn Stokes, Takashi Umeda, Ettore Vicari for
sending their results and to Frithjof Karsch, Peter Pereczky and Judah Unmuth-Yockey for careful reading
and comments on the manuscript.

\bibliography{bazavov_thermo_lat14.bbl}

\end{document}